\documentclass[12pt, draftclsnofoot, onecolumn]{IEEEtran}
\usepackage{amsmath,amssymb,amsfonts}
\usepackage{algorithmic}
\usepackage{array,color}
  \usepackage{bm}
\usepackage{float}
\usepackage{threeparttable}
\usepackage{color}
\usepackage{bbding}
\usepackage{subfig}
  \usepackage[font=small,labelsep=period]{caption}
\usepackage{array,color}
\usepackage{colortbl}
\usepackage{amsfonts,amssymb}
\usepackage{CJK}
\usepackage{booktabs}
\usepackage{multirow}
\usepackage{url}
\usepackage{stfloats}
\usepackage{diagbox}
\usepackage{amsmath}
\usepackage{autobreak}
\usepackage{cite}

\newtheorem{lemma}{Lemma}
\newtheorem{theorem}{Theorem}
\newtheorem{property}{Property}
\usepackage{subfloat}
\DeclareSubrefFormat{parens}{#1(#2)}

\usepackage{float}

\ifCLASSINFOpdf
   \usepackage[pdftex]{graphicx}
  \graphicspath{{../pdf/}{../jpeg/}}
  \DeclareGraphicsExtensions{.pdf,.jpeg,.png}
\else
  \usepackage[dvips]{graphicx}
  \graphicspath{{../eps/}}
  \DeclareGraphicsExtensions{.eps}
\fi
\begin{document}
% \title{A Secure Transmission System with \\ High Spectrum Efficiency Based on Faster-than-Nyquist and Deep Learning}
\title {For Intelligence and Higher Spectrum Efficiency: A Variable Packing Ratio Transmission System Based on Faster-than-Nyquist and Deep Learning}

%\author{\IEEEauthorblockN{Peiyang Song}
%\IEEEauthorblockA{State
%Key Laboratory of Integrated\\
%	 Service Networks \\
%	 Xidian University \\
%Xi'an, China\\
%Email: pysong@stu.xidian.edu.cn}
%}
%\\
%\IEEEauthorblockN{Qiang Li}
%\IEEEauthorblockA{State
%	Key Laboratory of Integrated\\
%	Service Networks\\
%	Xidian University \\
%	Xi'an, China \\
%	Email: qli\_4@stu.xidian.edu.cn}
%\and
%\IEEEauthorblockN{Nan Zhang}
%\IEEEauthorblockA{State
%	Key Laboratory of Integrated\\
%	Service Networks\\
%	Xidian University \\
%	Xi'an, China \\
%	Email: nzhang@xidian.edu.cn}
%\\
%\IEEEauthorblockN{Haiyang Ding}
%\IEEEauthorblockA{School of Information and Communications\\
%	National University of \\
%	Defense Technology \\
%	Xi'an, China \\
%	Email: dinghy2003@hotmail.com}
%\and
%\IEEEauthorblockN{Fengkui Gong}
%\IEEEauthorblockA{State
%Key Laboratory of Integrated\\
%	Service Networks\\
%	Xidian University \\
%	Xi'an, China \\
%Email: fkgong@xidian.edu.cn}}


\author{
	Peiyang Song\IEEEauthorrefmark{1}, \IEEEmembership{Student~Member,~IEEE},
	~Nan Zhang\IEEEauthorrefmark{1},
	~Lin Cai\IEEEauthorrefmark{2}, \IEEEmembership{Fellow,~IEEE}, \\
	~Guo Li\IEEEauthorrefmark{1}, \IEEEmembership{Member,~IEEE},
	~Tong Wu\IEEEauthorrefmark{3}, and 
	~Feng-kui Gong\IEEEauthorrefmark{1}, \IEEEmembership{Member,~IEEE}
	
	\IEEEauthorblockA{\IEEEauthorrefmark{1}State Key Laboratory of ISN, Xidian University, Xi'an, 710071, China}  \\
	\IEEEauthorblockA{\IEEEauthorrefmark{2} University of Victoria, BC V8W3P6, Canada}  \\
	\IEEEauthorblockA{\IEEEauthorrefmark{3}CAST-Xi’an Institute of Space Radio Technology, Xi’an 710071, China}  \\
	\IEEEauthorblockA{Email: {pysong@stu.xidian.edu.cn, nzhang@xidian.edu.cn, cai@ece.uvic.ca, gli@xidian.edu.cn, wut40@cast504.com, fkgong@xidian.edu.cn}} 
}


\maketitle

\begin{abstract}
	With the rapid development of various services in wireless communications, spectrum resource has become increasingly valuable. Faster than Nyquist (FTN) signaling, proposed in the 1970s, is a promising paradigm for improving spectrum utilization. 
This paper proposes the intelligent variable-packing-ratio (VPR)-based transmissions for high spectrum efficiency (SE) and security, respectively. Aided by deep learning (DL)-based estimation, the proposed scheme for high SE can achieve a higher capacity with negligible modification to existing communication paradigms (e.g., spectrum allocation or frame structure). Also, for VPR-based secure transmission, a dynamic generation scheme is proposed to produce randomly distributed positions to switch the packing ratio, which can effectively avoid detections and attacks.
	In addition, we propose a simplified DL-based packing ratio estimation for both of these two scenarios so that the receiver can estimate the packing ratio without any in-band or out-band control messages. Simulation results show that the proposed simplified estimation achieves nearly the same accuracy and convergence speed as the original multi-branch fully-connected structure with a complexity reduction of 20 folds.
	Finally, we derive the closed-form SE of the proposed VPR transmission under different channels. The numerical results validate the correctness of the derivation and demonstrate the SE gains of the VPR scheme beyond conventional Nyquist transmission. 

\end{abstract}

\IEEEpeerreviewmaketitle

\section{Introduction}
The last several decades have witnessed the rapid development of terrestrial wireless communications, including the widely concerned fifth-generation mobile communications (5G) and the increasing demands for data traffic by various communication services. 
%However, due to the limited coverage area and economic reasons, a large population is still excluded by the terrestrial communication networks. In recent years, satellite communications have attracted more attention in both academic and industrial fields for their wide coverage and the ability to provide seamless service for users located in remote areas (e.g., oceans, deserts and mountains).

Faster than Nyquist (FTN) signaling was firstly proposed in the 1970s by \emph{Bell Laboratories} and has been investigated and studied since the 2000s. It is promising to provide a higher symbol rate and spectrum efficiency (SE) in future terrestrial and satellite communications.
%, e.g., digital video broadcasting satellite second generation extension (DVB-S2X) \cite{li2020beyond}. 

In conventional Nyquist-criterion communications, the symbol duration must be set as $T>T_N=1/(2W)$ to guarantee the performance of the transmission system, where  $W$ is the transmission bandwidth. In such scenarios, the receiver can effectively recover the transmitted symbols from received ones benefiting from the strict orthogonality between different symbols. FTN signaling, in contrast, destroys the orthogonality and introduces unavoidable inter-symbol interference (ISI) by applying a smaller symbol duration $T<T_N$. It can improve the transmission rate, at the cost of higher complexity in the receiver to recover the transmitted symbols.

\begin{figure*}[!ht]
	\centering
	\includegraphics[width=0.6 \linewidth]{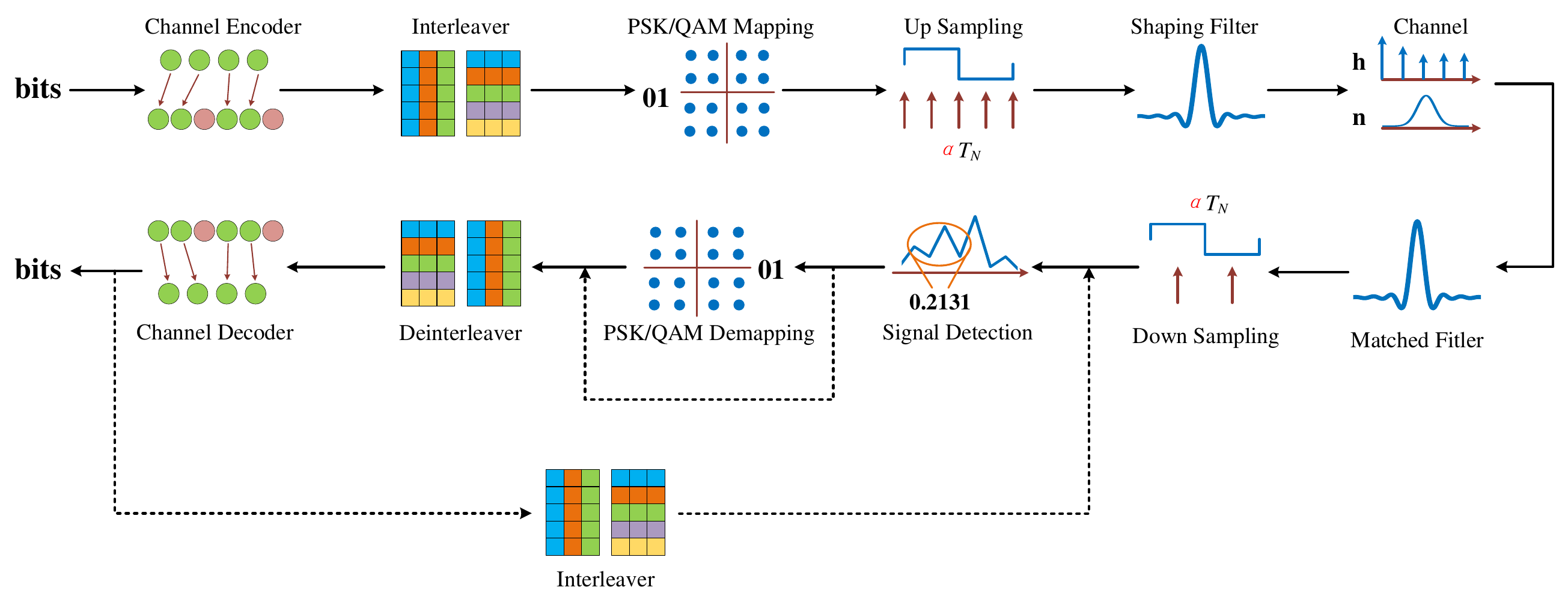}
	\caption{System model of conventional FTN signaling.}
	\label{fig:block}
\end{figure*}

 Mazo \cite{mazo1975faster} has proved that the FTN signaling can improve nearly 25\% rate than the conventional Nyquist-criterion communications in the additive white Gaussian noise (AWGN) channel without loss of bit error rate (BER) performance, which is known as \emph{the Mazo limit}.

 Many pieces of research have been conducted on the signal detection for FTN signaling, including time-domain \cite{liveris2003exploiting, anderson2009new, bedeer2017very, song2020dl, liu2021novel, petitpied2021circular, ibrahim2021novel} and frequency-domain \cite{sugiura2013frequency, ishihara2016frequency} algorithms. 
 %Among time-domain detections, \cite{liveris2003exploiting} and \cite{anderson2009new} employ the Viterbi algorithm, where FTN signaling is considered as a special type of convolutional codes. In \cite{bedeer2017very}, a very low-complexity symbol-by-symbol detection is developed. The combination of DL and FTN detection is studied in \cite{song2020dl} and \cite{liu2021novel} where DNN-based detection and sum-product detection are proposed respectively. \cite{petitpied2021circular} employs expectation propagation (EP) in the detection of FTN signaling. \cite{ibrahim2021novel} develops an FTN detector based on primal-dual predictor-corrector interior point method. Meanwhile, there are also a number of papers focusing on frequency-domain detection. References \cite{sugiura2013frequency} and \cite{sugiura2014frequency} apply the minimum mean square error (MMSE) criterion and propose frequency-domain equalizers for FTN signaling. Reference \cite{ishihara2016frequency} further considers channel estimation and develops an iterative detection algorithm.
Also, for sake of available high SE, some researchers attempt to merge FTN signaling with various conventional technologies such as frequency division multiplexing (FDM) \cite{rusek2009multistream, che2021multicarrier, ishihara2022reduced, ma2021generalized, anderson2006improving}, multiple input multiple output (MIMO) \cite{rusek2009existence, abebe2018ftn, yuhas2015capacity, mcguire2016faster}, multi-path fading channel \cite{wen2022joint, ishihara2017iterative, li2020joint}, etc. The comprehensive review of the latest study on FTN signaling can be found in \cite{ishihara2021evolution, zhou2019digital, fan2017faster}. Especially, \cite{rusek2009constrained} firstly derives the analytical-form capacity of FTN signaling, which inspires our work to extends it to the closed-form expressions and more scenarios.

The packing ratio is a key parameter that can directly affect the symbol rate and the strength of ISI. Conventional FTN signaling considers a fixed packing ratio which may not always achieve the maximum capacity during variable transmission conditions. A variable packing ratio (VPR)-based FTN signaling is a promising solution to this issue. Also, although the hopping roll-off factors \cite{wang2018hopping} have been successfully employed to improve the security of communications, the VPR-based secure transmission has not been studied yet. Last but not least, the success of deep learning (DL) in packing ratio estimation inspires us to develop intelligent VPR-based transmissions for the high SE and security.

The contribution of this paper can be summarized as follows.
\begin{itemize}
		\item We present an intelligent high SE VPR-based transmission based on FTN and DL. The transmitter can change the packing ratio based on specific conditions (e.g., channel state information (CSI) or cooperative strategy). No in-band or out-band control messages are required to notify the receiver of the packing ratio values, which means the scheme doesn't need to conduct a complex modification for existing communication paradigms (e.g., spectrum allocation and frame structure).
		\item We propose a VPR-based secure transmission, where the positions to change the packing ratio are secret and known only by the transmitter and the receiver. Also, no control messages are required since the receiver can infer the packing ratio with the DL-based simplified estimation and the information of positions.
		\item We propose a dynamic generation scheme for positions to change the packing ratio. With the measured CSI between the transmitter and the receiver, randomly distributed positions can be generated, which are secret to any other eavesdroppers.
		\item We propose a simplified DL-based packing ratio estimation, which achieves nearly the same performance as the original architecture while reducing the computing cost by 20 times.
		\item  We derive the closed-form expression of the capacity for the proposed VPR scheme in different channels and validate the theoretical results by Monte Carlo simulations. The derived capacities are also applicable to conventional FTN signaling.
		\item We conduct comprehensive evaluations and verify the SE gain between the proposed VPR-based and conventional Nyquist-criterion transmissions under different channels. Also, with the same SE, the BER degradations of the proposed VPR-based transmission over FTN signaling are demonstrated to be small enough.
	\end{itemize}

Herein, we give the definition of notations throughout the rest of the paper. Bold-face lower case letters (e.g. $\boldsymbol{x}$) are applied to denote column vectors. Light-face italic letters (e.g. $x$) denote scalers. $x_{i}$ is the $i$-th element of vector $\boldsymbol{x}$. $x(t)\ast y(t)$ denotes the convolution operation between $x(t)$ and $y(t)$. And $\lVert \boldsymbol{W}\rVert_0$ represents the number of non-zero items in matrix $\boldsymbol W$.

The rest of the paper is organized as follows. In Section \ref{sec:model}, we present the system model of FTN signaling. In Section \ref{sec:vpr}, the structure of the proposed VPR system is introduced. Section \ref{sec:dynamic} presents the proposed dynamic generation for positions of segments. And the simplified DL-based packing ratio estimation is presented in Section \ref{sec:simplified}. The capacity of the proposed VPR system under different channels is derived in Section \ref{sec:rate}. In Section \ref{sec:numerical}, comprehensive simulations are conducted to evaluate the performance and the complexity of the proposed VPR system and the DL-based estimation. Also, the derived capacity for the proposed VPR system is verified. Section \ref{sec:conclusion} concludes this paper.

\section{System Model of Conventional FTN Signaling} \label{sec:model}
This paper considers the complex-valued quadrature amplitude modulation (QAM) and AWGN channel. Fig. \ref{fig:block} illustrates the conventional architecture of FTN signaling.
In the transmitter, the signal that has passed through the shaping filter can be written as
\begin{equation}
s(t)=\sqrt{P_s}\sum_{k=-\infty}^{+\infty}x_{k}h(t-k\alpha T_{N}),
\end{equation}
where $P_s$ is the average power of the bandwidth signals, $x_k$ $(k=0,\pm 1, \pm 2, \cdots)$ is the $k$-th symbol and $\alpha$ ($0<\alpha \le 1$) is the symbol packing ratio. $h(t)$ is the function of shaping filters. Since the value of the filter function is 0 at every multiple of $T_N$, when $\alpha<1$, the filtered symbols are no longer orthogonal and become the weighted sum of several successive symbols.

Corresponding to the shaping filter, a filter with a conjugate structure named matched filter is employed in the receiver to maximize the received symbols' signal-to-noise ratio (SNR). The filtered symbols can be written as
\begin{equation}
y(t) = \left(s\left(t\right)+n\left(t\right)\right) \ast h(t) =\sqrt{E_{s}}\sum_{k=-\infty}^{+\infty}x_{k}g(t-k\alpha T_{N})+\widetilde{n}(t),
\end{equation}
where $g(t)=\int h(x)h(t-x)dx$, $\widetilde{n}(t)=\int n(x)h(t-x)dx$, and $n(t)$ is the Gaussian white noise.

Finally, the samples of the received symbols can be formulated as

%\begin{align} \label{eq:received_samples}
%y_{n}  & =\sqrt{E_{s}}\sum_{k=-\infty}^{+\infty}x_{k}g(n\alpha T_{N}-k\alpha T_{N})+\widetilde{n}(n\alpha T_N)\nonumber \\
%& =\underset{Interference \,from\,previous\,\,symbols}{\underbrace{\sqrt{E_{s}}\sum_{k=-\infty}^{n-1}x_{k}g\left(\left(n-k\right)\alpha T_{N}\right)}}+\sqrt{E_{s}} x_{n}g(0)\nonumber \\
%& \quad+\underset{Interference\,from\,upcoming\,\,symbols}{\underbrace{\sqrt{E_{s}}\sum_{k=n+1}^{+\infty}x_{k}g\left(\left(n-k\right)\alpha T_{N}\right)}}+\widetilde{n}\text{\ensuremath{\left(n\alpha T_N)\right)}}.
%\end{align}

\begin{align} \label{eq:received_samples}
y_{n} =\sqrt{E_{s}}\sum_{k=-\infty}^{n-1}x_{k}g\left(\left(n-k\right)\alpha T_{N}\right)+\sqrt{E_{s}} x_{n}g(0) +{\sqrt{E_{s}}\sum_{k=n+1}^{+\infty}x_{k}g\left(\left(n-k\right)\alpha T_{N}\right)}+\widetilde{n}\text{\ensuremath{\left(n\alpha T_N)\right)}}.
\end{align}

Different from the conventional Nyquist-criterion transmission system, each sampled symbol in FTN signaling contains both the expected symbol and the adjacent ones. Meanwhile, due to the non-orthogonality between different samples in the matched filter, the noise becomes colored noise. All these new features make it difficult to recover the original symbols in the FTN receiver.

\section{The Proposed Variable Packing Ratio Transmission System} \label{sec:vpr}

%As seen in Section \ref{sec:model}, in conventional FTN signaling, the packing ratio $\alpha$ is a key parameter that determines the tradeoff between transmission rate and the ISI strength. Inspired by the tradeoff, we develop VPR systems based on FTN and DL to achieve a higher SE and security.

\subsection{System Architecture}

\begin{figure}[ht]
	\centering
	\includegraphics[width=0.25 \linewidth]{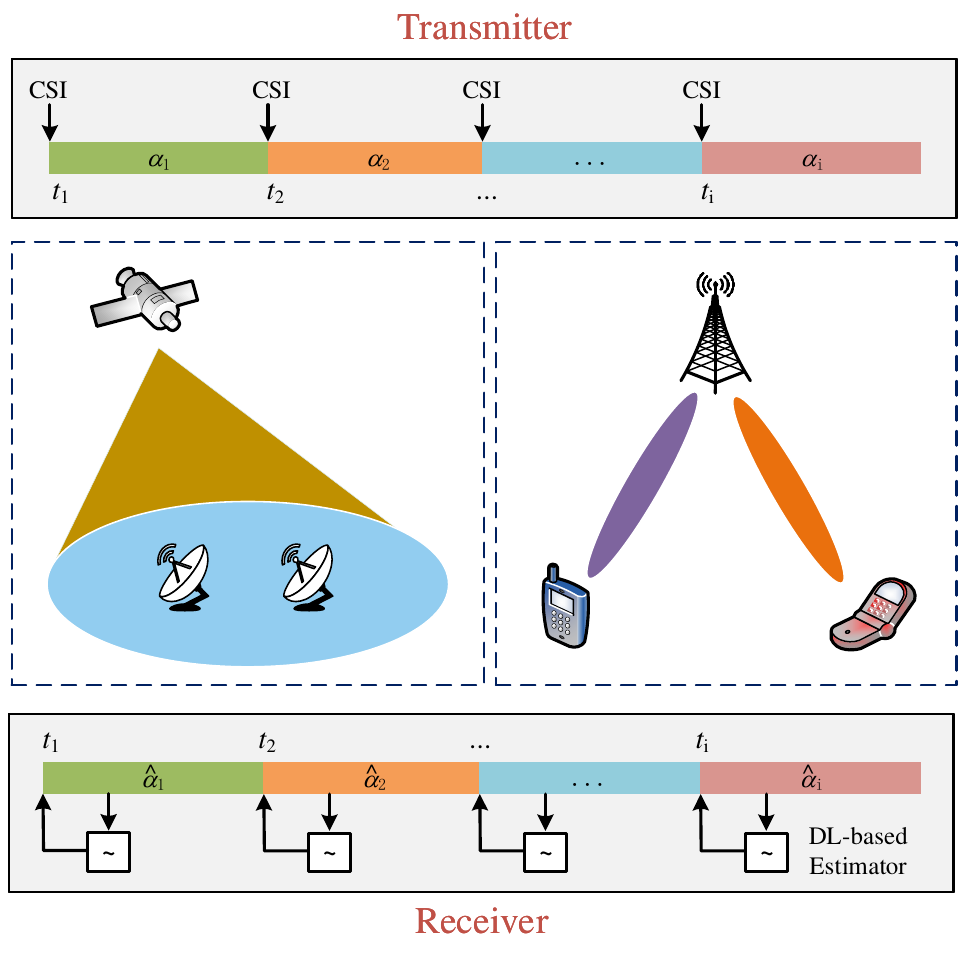}
	\caption{Architecture of the proposed VPR transmission scheme with CSI as the indicator to adjust the packing ratio.}
	\label{fig:test2}
\end{figure}
	
As shown by Fig. \ref{fig:test2}, in the proposed VPR transmission system, the transmitter changes the symbol packing ratio at every specific moment, which divides the transmitted symbols into different segments and results in individual transmission rates within each part. The determination of each packing ratio is based on CSI, cooperative target or other possible strategies.  Different from the conventional variable coding and modulation (VCM) schemes \cite{dvb_standard}, the receiver in the proposed VPR system does not need to know precisely the current symbol packing ratio. The only necessary knowledge is when the parameter changes, which can be appointed in advance. Then, the DL-based estimation will help the receiver infer the packing ratio within a short time.

There are two advantages to employ intelligent estimation for $\alpha$ in the receiver instead of directly sending it by the transmitter. Considering high SE, since no control message is required, the proposed VPR conducts negligible modification to the existing communication paradigms (e.g., spectrum allocation or frame structure).  Also, when considering security, if $\alpha$ is put into the frame head, the repeated specific modulation type and UW word will make it easy for the eavesdropper to locate and decode the information.

\subsection{VPR-based Transmission for High SE}

In this scenario, the packing ratio should be determined to balance the SE and the performance constraint (e.g., BER or ISI strength). For example, the SNR where $BER=10^{-3}$ can be employed as the threshold to select the packing ratio to achieve the maximum SE with acceptable BER performance, as demonstrated in Section \ref{sec:numerical}. Also, the signal-to-interference-plus-noise ratio (SINR) can be applied for the base station and satellite to adjust the packing ratio in non-orthogonal multiple access (NOMA) which divides users into pairs and the multi-beam satellite serving users within a certain area.

The positions when the packing ratio changes can be generated by the following approaches.

\begin{enumerate}
		\item \textbf{Fixed interval.} After the synchronization, the transmitter checks the transmission status after every fixed interval and decides whether to change the packing ratio. The receiver should carry out the estimation at the same positions.
		\item \textbf{Static storage.} A preset vector of starting positions is determined with the practical characteristics of the transmission environment and the requirement. 
		\item \textbf{Pilot or dedicated channel.} Without considering security, a public object (e.g., dedicated channel or pilot) can directly carry the information of packing ratio, at the expense of extra resources consumed and the modification of existing resources allocation.
	\end{enumerate}

\subsection{VPR-based Secure Transmission}
The VPR-based transmission is a promising paradigm to improve the security of communications. For one thing, the change of the symbol packing ratio only affects the baseband symbols and can not be caught by analysis of the frequency spectrum. For another, the blind estimation cannot indicate the accurate starting position. Once the eavesdropper employs a wrong symbol packing ratio, the sampled points will severely deviate from their correct positions, making it meaningless to detect the signals and further estimate the following symbol packing ratio.

In this scenario, the packing ratio should be employed randomly with the same probability to avoid possible detection and attack, as assumed for the roll-off factors in \cite{wang2018hopping}. The positions when the packing ratio changes can be generated by the following approaches.

\begin{enumerate}
	\item \textbf{Static storage.} A preset vector of starting positions should be stored in advance.  Although such assumptions have been widely employed \cite{wang2018hopping}, it suffers from the risk that the expected security will disappear once the information is stolen by the eavesdropper.
		
	\item \textbf{Dynamic generation.} A dynamic generation can effectively avoid the risk resulting from information leakage. In this paper, we utilize the fact that the CSI is known only by the two sides of communications and propose a dynamic scheme to generate a secret sequence of starting positions. The following section will give a detailed introduction on it.
\end{enumerate}

%As it is not the critical point of this paper, we just provide a preliminary analysis of it in the next part.

\section{The Proposed Dynamic Generation Scheme for Positions of Segments}  \label{sec:dynamic}

\subsection{The Architecture of the Dynamic Generation Scheme}
\begin{figure}[!ht]
		\begin{center}
			\includegraphics[width=0.4\textwidth]{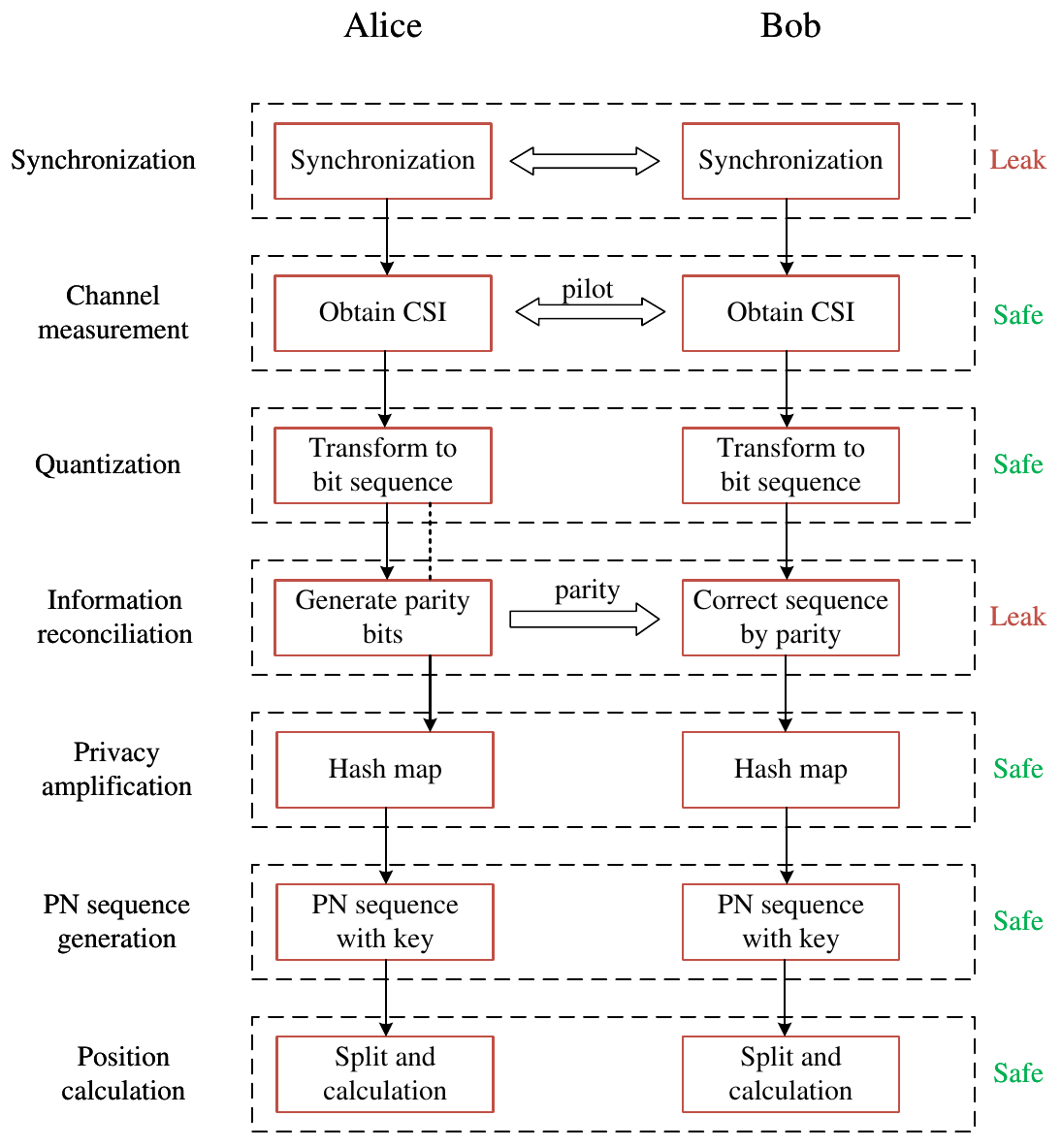}
		\end{center}
		\caption{The proposed dynamic generation scheme for the information of starting positions.}
		\label{fig:secrutiy_scheme}
\end{figure}

This section presents the proposed dynamic generation scheme for the starting positions of each segment where a new packing ratio is employed. The architecture of the scheme is shown in Fig. \ref{fig:secrutiy_scheme}. Alice, Bob, and Eve represent the transmitter, the receiver and the eavesdropper, respectively. The detailed steps are presented as follows.

\begin{itemize}
	\item \textbf{Channel measurement.} 
	In this stage, the two sides of communications send pilots to each other to measure the channel characteristics (e.g., signal intensity or channel response).

	\item \textbf{Quantization.}
	In this step, the transmitter and receiver quantize the measured information into a bit sequence by single \cite{aono2005wireless}, double \cite{mathur2008radio} or multiple-threshold \cite{patwari2009high} quantization.
	
	%todo 需要添加参考文献
	
	\item \textbf{Information reconciliation.}
	This step is conducted to correct the errors between the quantized bit sequence of the transceivers. For example, in Cascade-based information reconciliation \cite{brassard1993secret}, Alice splits the bits into segments and sends parity check information to Bob. Bob checks the parity states with corresponding pieces. Once the parity bits mismatch, a binary search is conducted by changing as few bits as possible to satisfy the parity requirement.
			
	\item \textbf{Privacy amplification.}
	Generally speaking, there always exists some information that leaks to Eve in the information reconciliation stage. By mapping the quantized information into a new bit sequence (i.e., the secret key) with the hash function (e.g., the message-digest algorithm 5 (MD5)), the risk resulting from the partly leaked information can be eliminated.
			
	\item \textbf{Pseudo{-}noise (PN) sequence generation.}
	In this step, the secret key is employed as the seed for the PN generator, which can produce random and unrelated bits within its period.
	
	\item \textbf{Position calculation.}
	The offsets can be easily obtained by splitting the PN sequence into segments with the same length and transforming them into signed integers. Then, the starting positions can be calculated by adding them to the original positions with the fixed step. 
\end{itemize}

\subsection{Performance Analysis for the Dynamic Generation Scheme}
\begin{enumerate}
	\item \textbf{Security.}
	The CSI between Alice and Bob can only be measured by both of them, Eve cannot obtain it even if the pilot is eavesdropped. Except the information reconciliation, the other stages are safe since they are executed internally without any information exchanged.

	Although limited information may be leaked by the parity sent to Bob, the hash mapping operation enhances the system security by mapping the information bits to a new bit sequence (i.e., the secret key) which cannot be inferred by partial original information. 
	
	So, throughout this paper, the private key, as well as the generated positions, are considered to be secret and cannot be obtained by Eve.
	
	\item \textbf{Randomness.}
	In fact, this issue has been studied in the physical layer security field. The national institute of standards and technology (NIST) test \cite{rukhin2001statistical} is widely employed to measure the randomness of the generated secret key. There have been many existing CSI-based key generation schemes that pass the NIST test.
	
	\item \textbf{Robustness.}
	Another import metric is the robustness, which means the proposed scheme can work well under various scenarios and guarantee enough randomness.
	
	The work about this issue can also be found in the existing literature. For example, the secret key generation under different transmission, e.g., MIMO \cite{madiseh2012applying} and multi-carrier communications \cite{aldaghri2020physical}, have been widely studied. The solution in different channels (e.g., multi-path channel \cite{sayeed2008secure} and even the static channel \cite{patwari2009high, aldaghri2020physical}) have also been presented.
	
	\item \textbf{Period.}
	The PN generator has a period of $2^{N_r}$, where $N_r$ is the number of the registers (i.e., the number of bits in the PN generator's state). During the period, the generated bits has an excellent autocorrelation feature which achieves nearly an impulse function. When the generated bits are divided by step $N_s$, the period of the calculated offset is $\rm lcm(2^N_r, N_s)$, where ${\rm lcm}(a, b)$ means the least common multiple of $a$ and $b$.
\end{enumerate}

\section{A Simplified Symbol Packing Ratio Estimation for FTN Signaling}  \label{sec:simplified}
\begin{figure}[ht]
	\centering
	\includegraphics[width=0.35\linewidth]{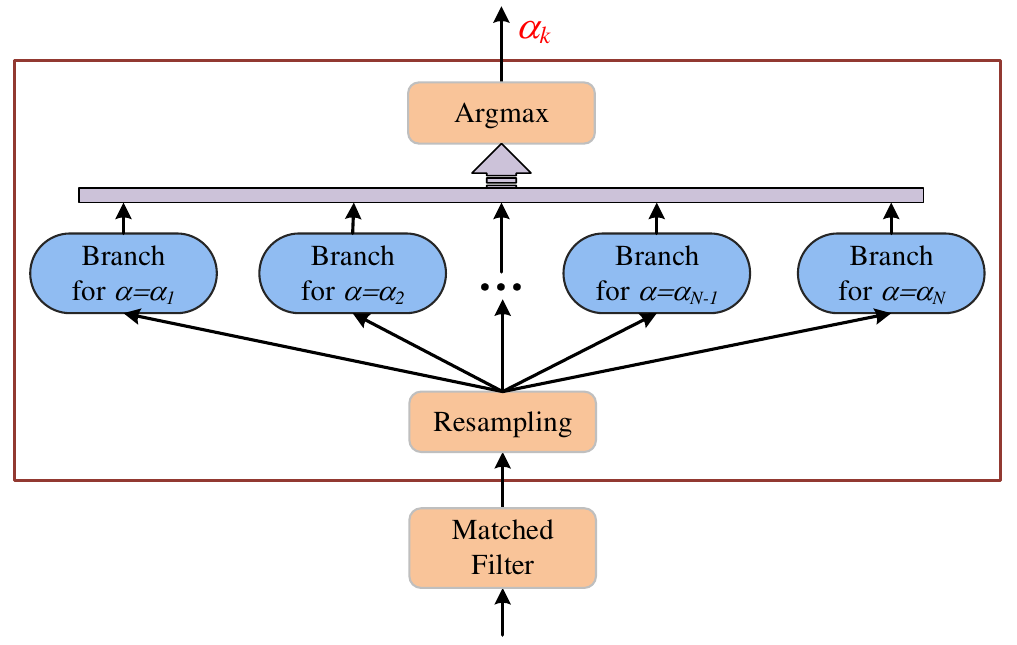}
	\caption{Structure of the symbol packing ratio estimation employed in the proposed system.}
	\label{fig:totalestimation}
\end{figure}

In this part, we present a simplified symbol packing ratio estimation for FTN signaling. Fig. \ref{fig:totalestimation} illustrates the complete architecture of the proposed estimation. The symbols that have passed through the matched filter and then been sampled are applied as the input of several analysis models. The main task of the analysis for $\alpha_k$ is to decide whether $\alpha=\alpha_k$, where $\alpha$ is the correct symbol packing ratio employed by the transmitter.

\begin{figure}[ht]
	\centering
	\includegraphics[width=0.4\linewidth]{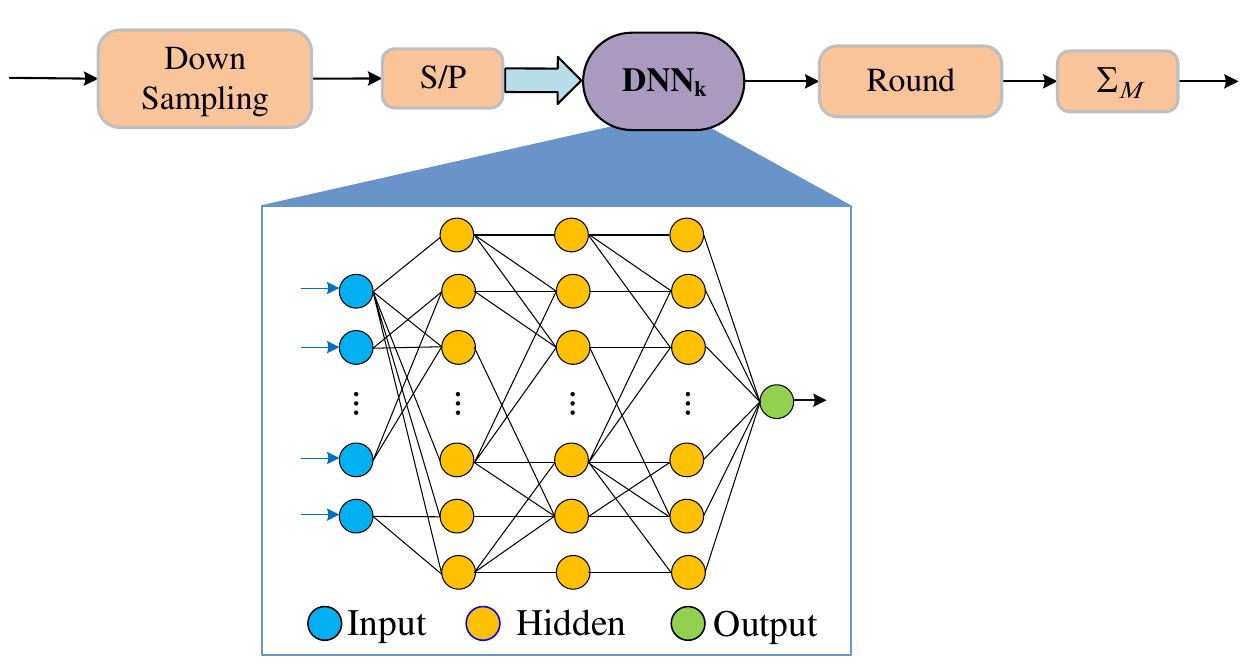}
	\caption{Structure of the analysis for $\alpha_k$ in the proposed simplified symbol packing ratio estimation.}
	\label{fig:subblock}
\end{figure}

 Fig. \ref{fig:subblock} shows the detailed structure of the branch for analyzing whether $\alpha=\alpha_k$. Firstly, the input symbols are down-sampled by the shared knowledge of starting position and interval $\alpha_i T_N$. Then, through serial-parallel conversion (S/P), the sampled serial symbols are reformed and fed into the deep neural network (DNN) \cite{goodfellow2016deep}. The output of DNN can be regarded as the probability of $\alpha_A=\alpha_k$ and will be transformed into integer 0 (false) or 1 (true). And finally, the number of true decisions during a specific time will be counted.

The DNN we employed in Fig. \ref{fig:subblock} contains an input layer, three hidden layers and an output layer. Each hidden layer is a sparsely connected layer with ReLU as its activation function. The system function of the DNN can be written as

\begin{equation}\label{eq:dnn}
\mathbf{y}=g_{4}\left(f\left(g_{3}\left(f\left(g_{2}\left( f\left(g_{1}(\boldsymbol{x})\right)\right)\right)\right)\right)\right. ,
\end{equation}
where $f(\boldsymbol{x})_{i}=\max \left(x_{i}, 0\right)$ is the item-wise ReLU function to vector $\boldsymbol{x}$. $g_{i}(\boldsymbol{x})=\boldsymbol{W}_{i} \boldsymbol{x}+\boldsymbol{b}_{i}$, where $\boldsymbol{W}_i$ and $\boldsymbol{b}_{i}$ are the weight matrix and bias vector in the $i$-th layer of the DNN.

Benefiting from that the information of starting position for each transmission segment with a new $\alpha$ is known by both the transceiver, the receiver does not need to divide the signal into several streams \cite{song2019blind} to avoid the sampling offset. So, the multiplexer (MUX), the demultiplexer (DEMUX) and the decision model in the original structure can be removed.

Meanwhile, we focus on the simplification of DNN. The main idea is to reduce the amounts of items in the weight matrices. Here, we employ an iterative strategy. After the model is well trained, we remove the items in $\boldsymbol{W}_i$ that are small enough and then train the remaining network. The process will be iteratively carried out until the target sparsity ratio is reached.

%which contains two stage named \emph{training} and \emph{pruning}.  In the training stage, we train the network following the common approaches. And then, in the pruning stage, we compare the network with the certain threshold $v_t$ where $t$ is the iteration index. We set the nodes to $0$ and do not train them again when they are less than $v_t$. After that, we return to the previous stage to train the remaining network and iteratively carry out these processes until the designed iteration times.

\section {Spectrum Efficiency of Proposed VPR System in Different Channels} \label{sec:rate}
Generally speaking, the proposed VPR-based system can work well on various channels as long as FTN signaling are applicable. Here, we consider the AWGN, Rayleigh \cite{rice1944mathematical} and Nakagami-m \cite{nakagami1960m} channels as the examples. The Rayleigh and Nakagami-m channels are typical fading channels that were first studied in 1940s and 1960s, repectively.

\subsection{Theoretical SE of VPR System in AWGN Channel}
In the receiver, $\alpha$ can be easily obtained with the help of blind estimation and the exact starting position. So, $\alpha$ can be regarded as the shared information between the transmitter and the receiver. And the transmission can be considered to be a conventional FTN signaling. With power $\sigma_s^2=P_s\alpha T_N$ for the transmitted signal, the analytical-form capacity of FTN signaling can be been formulated by Rusek as \cite{rusek2009constrained}
\begin{equation}\label{eq:FTN_rate}
R_A(\alpha)=\frac{1}{2 \pi \alpha T_N} \int_{0}^{\pi} \log _{2}\left(1+\frac{2 \sigma_{s}^{2}}{N_{B}} H(\alpha, \omega)\right) \rm d \omega,
\end{equation}
where $N_B/2$ is the power spectrum density of the Gaussian noise in the AWGN channel. And $H(\alpha, \omega)$ is defined by
\begin{equation}\label{eq:H_omega}
H(\alpha, \omega)=\frac{1}{\alpha T_N} \sum_{k=-\infty}^{\infty}\left|G\left(\frac{\omega}{2 \pi \alpha T_N}+\frac{k}{\alpha T_N}\right)\right|^{2},
\end{equation}
where $G(f)$ is the Fourier transform of $h(t)$. $|G(f)|^2$ can be expressed as \cite{cubukcu2012root}

\begin{equation} \label{eq:square_g}
|G\left( f \right)|^2 =
		\begin{cases}
			T_N, & |f|\in \left[ 0,\frac{1-\beta}{2T_N} \right] \\
			\frac{T_N}{2}\left\{ 1+\cos \left[ \frac{\pi T_N}{\beta}\left( \left| f \right|-\frac{1-\beta}{2T_N} \right) \right] \right\} , & |f|\in \left[ \frac{1-\beta}{2T_N},\frac{1+\beta}{2T_N} \right]\\
		0, & |f|\in \left[ \frac{1+\beta}{2T_N}, +\infty \right]\\
	
\end{cases}.
\end{equation}

Here, we define three bound functions $b_1\left( \alpha  \right)={\alpha \pi (1 - \beta )}$, ${b_2}\left( \alpha  \right) = 2\pi  - \alpha \pi \left( {1 + \beta } \right)$ and ${b_3}\left( {\alpha} \right) = \alpha \pi \left(1+\beta\right)$. And the following conclusion can be derived.

\begin{lemma} \label{lemma_1}
\rm When $\omega\in[0, \pi]$, for any $k\ge1$, it always holds that
\begin{equation} \label{eq:lemma1}
G^2\left( \frac{\omega +2k\pi}{2\pi \alpha T_N} \right) =0.
\end{equation}

\end{lemma}

\begin{IEEEproof}
	Here, we firstly assume that for any $k\ge 1$, it holds that
	
	\begin{equation} \label{eq:g_f1}
		\frac{\omega +2\pi k}{2\pi \alpha T_N}\ge \frac{1+\beta}{2T_N}.
	\end{equation}
	
	Since that $\omega\in[0, \pi]$, \eqref{eq:g_f1} can be proved by
	
	\begin{equation}
		\label{eq:g_f11}
			\frac{\omega +2\pi k}{2\pi \alpha T_N}\ge \frac{1+\beta}{2T_N}
			\Leftarrow  \frac{2\pi k}{2\pi \alpha T_N}\ge \frac{1+\beta}{2T_N}
			\Leftrightarrow  \frac{k}{\alpha}\ge \frac{1+\beta}{2}
			\Leftrightarrow  2k\ge \alpha (1+\beta ) .
	\end{equation}
	
	Considering that $k>1$ and $0\le\alpha,\beta\le 1$, \eqref{eq:g_f11} can be further obtained by
	
	\begin{equation} \label{eq:g_f111}
		2k\ge \alpha (1+\beta )\Leftarrow 2\ge \alpha \left( 1+\beta \right) \Leftarrow 2\ge 2.
	\end{equation}
	
	It's obviously that $2>2$ always holds. So, the assumption \eqref{eq:g_f1} is proved. And finally, \textit{Lemma} \ref{lemma_1} can be proved by combining \eqref{eq:square_g} with \eqref{eq:g_f1}.
\end{IEEEproof}

\begin{lemma} \label{lemma_2}
When $\omega\in[0, \pi]$, for any $k\le-2$, it always holds that
\begin{equation} \label{eq:lemma2}
	G^2\left( \frac{\omega +2k\pi}{2\pi \alpha T_N} \right) =0.
\end{equation}
\end{lemma}

\begin{IEEEproof}
	Firstly, we assume that for any $k\le-2$, it holds that
	\begin{equation} \label{eq:g_f2}
	\frac{\omega +2\pi k}{2\pi \alpha T_N}\le  -\frac{1+\beta}{2T_N}	.
	\end{equation}
	
	Since that $\omega\in[0, \pi]$,  \eqref{eq:g_f2} can be proved by
	\begin{equation} \label{eq:g_f21}
		\begin{aligned}
			\frac{\omega +2\pi k}{2\pi \alpha T_N}\le -\frac{1+\beta}{2T_N}\Leftarrow  \frac{\pi +2\pi k}{2\pi \alpha T_N}\le -\frac{1+\beta}{2T_N}
			\Leftrightarrow  2k\le -\alpha \left( 1+\beta \right) -1 .
		\end{aligned}
	\end{equation}
	
	Considering that $k\le-2$ and $0\le\alpha, \beta \le 1$, \eqref{eq:g_f21} can be obtained by
	\begin{equation} \label{eq:g_f211}
		2k\le -\alpha \left( 1+\beta \right) -1\Leftarrow -4\le -\alpha \left( 1+\beta \right) -1 \Leftarrow -4\le -3 .
	\end{equation}
	
	It's obviously that $-4\le-3$ always holds. So, \eqref{eq:g_f2} is proved. And finally, \textit{Lemma} \ref{lemma_2} can be proved by combining \eqref{eq:square_g} with \eqref{eq:g_f2}.
\end{IEEEproof}

\begin{theorem}  \label{theorem_1}
	For $\omega \in [0, \pi]$, $H(\alpha, \omega)$ can be expressed as
	\begin{equation}
		H(\alpha, \omega)=\frac{1}{\alpha T_N}\;\left( G^2\left( \frac{\omega}{2\pi \alpha \mathrm{T}_N} \right) +G^2\left( \frac{\omega -2\pi}{2\pi \alpha \mathrm{T}_N} \right) \; \right) .
	\end{equation}
\end{theorem}

\begin{IEEEproof}
	The theorem can be proved by combining \textit{Lemma} \ref{lemma_1}, \textit{Lemma} \ref{lemma_2} and \eqref{eq:H_omega}.
\end{IEEEproof}

\begin{lemma}  \label{lemma_3}
	\rm For $\omega\in\left[0, b_1(\alpha)\right)$, it always holds that
	\begin{equation}
		H(\alpha, \omega )=\frac{1}{\alpha}.
	\end{equation}
\end{lemma}

\begin{IEEEproof}
	Since that $\omega\in[0, b_1(\alpha))$, it can be obtained that
	\begin{equation}  \label{eq:g_3}
		0\le \frac{\omega}{2\pi \alpha T_N}\le \frac{b_1(\alpha)}{2\pi \alpha T_N}=\frac{1-\beta}{2T_N}	.
	\end{equation}
	
	Considering \eqref{eq:square_g} and \eqref{eq:g_3}, for $\omega\in[0, b_1(\alpha))$, it can be derived that
	\begin{equation} \label{eq:g_31}
		G^2\left( \frac{\omega}{2\pi \alpha T_N} \right) =\frac{1}{\alpha}	.
	\end{equation}
	
	Also, since $\omega\in[0,b_1(\alpha)$, it can be obtained that
	\begin{equation}  \label{eq:g_311}
		\frac{\omega -2\pi}{2\pi \alpha T_N}\le \frac{b_1(\alpha )-2\pi}{2\pi \alpha T_N}=\frac{\alpha \pi \left( 1-\beta \right) -2\pi}{2\pi \alpha T_N}.
	\end{equation}
	
	Considering that $0\le\alpha, \beta\le1$, \eqref{eq:g_311} can be further written as
	\begin{equation} \label{eq:g_3111}
		\frac{\omega-2\pi}{2\pi\alpha T_N} \le\frac{\alpha \pi \left( 1-\beta \right) -2\pi}{2\pi \alpha T_N}\le \frac{\alpha \pi \left( 1-\beta \right) -2\alpha \pi}{2\pi \alpha T_N}=-\frac{1+\beta}{2T_N}	.
	\end{equation}
	
	Considering \eqref{eq:square_g} and \eqref{eq:g_3111}, it can be derived that
	\begin{equation} \label{eq:g_31111}
		G^2\left( \frac{\omega -2\pi}{2\pi \alpha T_N} \right) =0.
	\end{equation}

	Finally, \textit{Lemma} \ref{lemma_3} can be proved by combining \textit{Theorem} \ref{theorem_1}, \eqref{eq:g_31} and \eqref{eq:g_31111}.	
\end{IEEEproof}

\begin{lemma}  \label{lemma_4}
	For $\omega\in[b_3(\alpha),\pi]$ and $b_2(\alpha)\ge\pi$, it always holds that
	\begin{equation}
		G^2\left( \frac{\omega}{2\pi \alpha T_N} \right) =0.
	\end{equation}

\end{lemma}

\begin{IEEEproof}
	Firstly, it can be obtained by $b_2(\alpha)\ge\pi$ that
	\begin{equation}
		b_2\left( \alpha \right) =2\pi -\alpha \pi \left( 1+\beta \right) \ge \pi 
		\\
		\Leftrightarrow \alpha \le \frac{1}{1+\beta} .	
	\end{equation}
	
	Then, since that $\omega\in[b_3(\alpha), \pi]$, it can be obtained that
	\begin{equation}  \label{eq:g_411}
		\frac{\omega -2\pi}{2\pi \alpha T_N}\le \frac{\pi -2\pi}{2\pi \alpha T_N}=-\frac{1}{2\alpha T_N}\le -\frac{1+\beta}{2T_N}.
	\end{equation}
	
	Considering \eqref{eq:square_g} and \eqref{eq:g_411}, it can be derived that
	\begin{equation} \label{eq:g_4111}
		G^2\left( \frac{\omega -2\pi}{2\pi \alpha T_N} \right) =0  	.
	\end{equation}

	% todo 这里需要加上限定条件
	Also, since that $\omega\in[b_3(\alpha), \pi]$, it can be obtained that
	\begin{equation}  \label{eq:g_41111}
		\frac{\omega}{2\pi \alpha T_N}\ge \frac{\alpha \pi \left( 1+\beta \right)}{2\pi \alpha T_N}=\frac{1+\beta}{2T_N}.	
	\end{equation}
	
	Considering \eqref{eq:square_g} and \eqref{eq:g_41111}, it can be derived that
	\begin{equation}  \label{eq:g_411111}
		G^2\left( \frac{\omega}{2\pi \alpha T_N} \right) =0.	
	\end{equation}
 
	Finally, \textit{Lemma} \ref{lemma_4} can be proved by combining \textit{Theorem} \ref{theorem_1}, \eqref{eq:g_4111} and \eqref{eq:g_411111}.	
\end{IEEEproof}

By combining \textit{Theorem} \ref{theorem_1}, \textit{Lemma} \ref{lemma_3} and \textit{Lemma} \ref{lemma_4}, $H(\alpha, \omega)$ can be expressed as

\begin{equation}
	H\left( \alpha ,\omega \right) =\begin{cases}
		H_1\left( \alpha ,\omega \right) ,&		b_2(\alpha )<\pi\\
		H_2\left( \alpha ,\omega \right) ,&		b_2\left( \alpha \right) \ge \pi\\
	\end{cases},	
\end{equation}
where
\begin{equation}
	H_1\left( \alpha ,\omega \right) =\begin{cases}
\frac{1}{\alpha},&		\omega \in \left[ 0,b_1\left( \alpha \right) \right)\\
\frac{1}{\alpha T_N}\;\left( G^2\left( \frac{\omega}{2\pi \alpha \mathrm{T}_N} \right) +G^2\left( \frac{\omega -2\pi}{2\pi \alpha \mathrm{T}_N} \right) \; \right) ,&		\omega \in \left[ b_1\left( \alpha \right) ,\pi \right]\\
\end{cases} ,
\end{equation}
%and
\begin{equation}
	H_2\left( \alpha ,\omega \right) =\begin{cases}
	\frac{1}{\alpha},&		\omega \in \left[ 0,b_1\left( \alpha \right) \right)\\
	\frac{1}{\alpha T_N}\;\left( G^2\left( \frac{\omega}{2\pi \alpha \mathrm{T}_N} \right) \; \right) ,&		\omega \in \left[ b_1\left( \alpha \right) ,b_3\left( \alpha \right) \right)\\
	0,&		\omega \in \left[ b_3\left( \alpha \right) ,\pi \right]\\
\end{cases}	.
\end{equation}
	
With the system bandwidth that can be calculated by $W=1/(2T_N)=W_T/(1+\beta)$, where $W_T$ is the total bandwidth of the channel, SE of FTN signaling can be written as
\begin{equation}\label{eq:FTN_cap}
C_A(\alpha )=\frac{1}{\pi \alpha (1+\beta )}\underset{C_B\left( \alpha \right)}{\underbrace{\int_0^{\pi}{\log _2}\left( 1+\frac{2\sigma _{s}^{2}}{N_B}H(\alpha ,\omega ) \right) \mathrm{d\omega}}}.
\end{equation}

 Then, we split $C_B\left(\alpha\right)$ into several subsection integral and calculate them respectively. For $\omega \in \left[0, b_1\left(\alpha\right)\right)$, the integral can be expressed as
\begin{equation}
\begin{split}
{C_{1}}\left( \alpha  \right) = \int_0^{{b_1}\left( \alpha  \right)} {{{\log }_2}\left( {1 + \frac{{2\sigma _s^2}}{{{N_B}}}H(\alpha ,\omega )} \right)} {\rm{d}}\omega   \vspace{1em} =\alpha \pi \left( {1 - \beta } \right){\log _2}\left( {1 + \frac{{2\sigma _s^2}}{{{\alpha N_B}}}} \right)
\end{split} .
\end{equation}

\begin{theorem} \label{theorem_3}
(\textit{Chebyshev-Gauss Quadrature Rule})  For a given function $f(x)$, its integration between -1 and 1 can be approximated as \cite{abramowitz1964handbook}

	\begin{equation}
		\int_{-1}^1{\frac{f\left( x \right)}{\sqrt{1-x^2}} dx\approx \sum_{i=1}^n{w_if\left( \xi_i \right)}},
	\end{equation}
	where $\xi_i = \cos \left(\frac{2i-1}{2n}\pi \right)$ and $w_i = \frac{\pi}{n}$.
\end{theorem}

\begin{theorem}  \label{theorem_4}
	\rm For given function $f(x)$, its integration between $a$ and $b$ can be approximated as
	
	\begin{equation}
		\int_a^b{f}(x)\,dx\approx \frac{b-a}{2}\sum_{i=1}^n{w_i}\sqrt{1-\xi_{i}^{2}}f\left( \frac{b-a}{2}\xi+\frac{b+a}{2} \right) ,
	\end{equation}
	where the values of $\xi_i$ and $w_i$ are the same as those in \textit{Theorem} \ref{theorem_3}.
\end{theorem}

\begin{IEEEproof}
Here, we set
$
	x=\frac{b-a}{2}\xi +\frac{b+a}{2}
$. Then, the integration of $f(x)$ can be rewritten as
\begin{equation} \label{eq:g_511}
	\begin{aligned}
		\int_a^b{f\left( x \right) dx}&=\int_{arg_{\xi}\left( a \right)}^{arg_{\xi}\left( b \right)}{f\left( \frac{b-a}{2}\xi +\frac{b+a}{2} \right) d\left[ \frac{b-a}{2}\xi +\frac{b+a}{2} \right]} \\
		&=\frac{b-a}{2}\int_{-1}^1{f\left( \frac{b-a}{2}\xi +\frac{b+a}{2} \right) d\xi}
		=\frac{b-a}{2}\int_{-1}^1{\frac{\sqrt{1-x^2}\psi \left( \xi \right)}{\sqrt{1-x^2}}d\xi} ,\\
	\end{aligned}
\end{equation}

where $\psi (\xi)$ is defined as
$
	\psi \left( \xi \right) =f\left( \frac{b-a}{2}\xi +\frac{b+a}{2} \right)
$.

Considering \textit{Theorem} \ref{theorem_3}, \eqref{eq:g_511} can be written as
\begin{equation}
	\begin{aligned}
		\int_a^b{f\left( x \right) dx}&\approx \frac{b-a}{2}\sum_{i=1}^n{w_i\sqrt{1-{\xi _i}^2}\psi \left( \xi_i \right)}
		=\frac{b-a}{2}\sum_{i=1}^n{w_i\sqrt{1-{\xi _i}^2}f\left( \frac{b-a}{2}\xi _i+\frac{b+a}{2} \right)} . \\
	\end{aligned}	
\end{equation}	
\end{IEEEproof}

According to the \textit{Theorem} \ref{theorem_4}, for $\omega \in \left[b_1\left(\alpha\right), \pi\right]$, the integral can be written as
\begin{equation}
C_2\left( \alpha \right) =\int_{b_1\left( \alpha \right)}^{\pi}{\log _2\left( 1+\frac{2\sigma _{s}^{2}}{N_B}H(\alpha ,\omega ) \right) d\omega}\approx A_1\sum_{i=1}^N{m_i\sqrt{1-\omega _{1i}^{2}}}\log _2\left( 1+\frac{2\sigma _{s}^{2}}{N_B}H(\alpha ,\omega _{1i}) \right)  ,
\end{equation}
where
$
A_1 = \frac{{\pi \left[ {1{\rm{ + }}\alpha {\mkern 1mu} \left( {\beta  - 1} \right)} \right]{\mkern 1mu} }}{2}
$, $
m_i = \frac{{\pi \left| {\sin \left( {\frac{{\pi \left( {2i - 1} \right)}}{{2N}}} \right)} \right|}}{N}
$
and
$
{\omega _{1i}} = \frac{\pi }{2}\left\{ {\left[ {1 + \alpha \left( {\beta  - 1} \right)} \right]\cos \left( {\frac{{\pi (2i - 1)}}{{2n}}} \right) + {1 - \alpha \left( {\beta  - 1} \right)}} \right\}
$.

Similarly, for $\omega \in \left[b_1\left(\alpha\right), b_3\left(\alpha\right)\right)$, the integral can be written as
\begin{equation}
C_3\left( \alpha \right) =\int_{b_1(\alpha )}^{b_3(\alpha )}{\mathrm{l}_2}\left( 1+\frac{2\sigma _{s}^{2}}{N_B}H(\alpha ,\omega ) \right) d\omega \approx A_2\sum_{i=1}^N{m_i}\sqrt{1-\omega _{2i}^{2}}\log _2\left( 1+\frac{2\sigma _{s}^{2}}{N_B}H(\alpha ,\omega _{2i}) \right),
\end{equation}
where 
$
A_2=\pi \alpha \beta	
$
and
$
{\omega _{2i}} = \pi {\mkern 1mu} \alpha \left( {1 + \beta {\mkern 1mu} {\rm{cos}}\left( {\frac{{\pi {\mkern 1mu} \left( {2{\mkern 1mu} {i} - 1} \right)}}{{2{\mkern 1mu} n}}} \right)} \right) .
$

For the convenience of implementation, the set of available $\alpha$ values is usually finite. Finally, for a specific $\alpha$ value, SE of the proposed VPR system in the AWGN channel can be written as
\begin{equation}
C_A\left( \alpha \right) =\left\{ \begin{matrix}
	\frac{1}{\alpha \pi \left( 1+\beta \right)}\left( C_1\left( \alpha \right) +C_2\left( \alpha \right) \right) ,&		b_2\left( \alpha \right) <\pi\\
	\frac{1}{\alpha \pi \left( 1+\beta \right)}\left( C_1\left( \alpha \right) +C_3\left( \alpha \right) \right) ,&		b_2\left( \alpha \right) \ge \pi\\
\end{matrix} \right..
\end{equation}

To avoid the possible detection and attack when the VPR system is employed to improve the security, every $\alpha$ is preferred to be applied with the same probability, just as the roll-off factor in \cite{wang2018hopping}. So, for the proposed VPR system, the average SE in such a scenario can be written as
 \begin{equation}\label{eq:awgn_average}
 C_A^\prime = \frac{1}{N_{\alpha}}\sum_{i=1}^{N_{\alpha}}C_A(\alpha_i),
 \end{equation}
 where $\alpha_i$ ($i=1,2\cdots, N_\alpha$) is the $i$-th symbol packing ratio that is employed in the transmission.
 
 \subsection{Theoretical SE of VPR System in Rayleigh Channel}
 For the Rayleigh and Nakagami-m channel, the channel gain is considered and can be regarded as a constant during every data block in this paper. So, the power of the signal in the receiver with channel gain $h$ can be written as
 \begin{equation}
 \sigma_{s^\prime}^2\left(h\right)=h^2P_s\alpha T_N .
 \end{equation}
 
 The capacity of FTN signaling with specific $h$ can be obtained as
 \begin{equation}
 	R^\prime(\alpha)=\frac{1}{2 \pi \alpha T_N} \int_{0}^{\pi} \log _{2}\left(1+\frac{2 \sigma_{s^\prime}^{2}\left(h\right)}{N_{B}} H(\alpha, \omega)\right) \rm d \omega .
 \end{equation}

 Considering that $h$ is a random variable, the mean SE of FTN signaling with packing ratio $\alpha$ in Rayleigh channel can be formulated as 
 \begin{equation} \label{eq:ray_aver_rate}
 \begin{split}
 	C_R(\alpha) = \frac{1}{{\pi \alpha \left(1+\beta\right)}} \int_0^\pi  {\underbrace {\int_0^{ + \infty } {f_R(h) \cdot {\rm{lo}}{{\rm{g}}_2}\left( {1 + \frac{{2{h^2}P_s\alpha {T_N}}}{{{N_B}}}H(\alpha ,\omega )} \right){{\rm d} h}} }_{{C_{i1}}(\alpha, \omega)}} {\rm{d\omega}} 
 \end{split} ,
 \end{equation}
 where $f_R(h)$ is the probability density function (PDF) of $h$, which can be written as \cite{rice1944mathematical}
\begin{equation}
	f_R(h) = \frac{h}{\sigma^2} e^{-\frac{h^2}{2\sigma^2}} ,
\end{equation}
where $\sigma^2$ is the power parameter. Then, by applying $C_{o1}(\alpha, \omega)=2P_s\alpha T_NH\left(\alpha, \omega\right)/N_B$, $C_{i1}\left(\alpha, \omega\right)$, which has been defined in (\ref{eq:ray_aver_rate}), can be written as
\begin{equation}
\begin{split}
& {C_{i1}}(\alpha ,\omega ) = \int_0^{ + \infty } {{\rm{ - lo}}{{\rm{g}}_2}\left( {1 + {C_{o1}(\alpha ,\omega)}{h^2}} \right)\left( { - \frac{h}{{{\sigma ^2}}}{e^{ - \frac{{{h^2}}}{{2{\sigma ^2}}}}}} \right){\rm{d}}h}.
\end{split}
\end{equation}

By extracting the integral items as $F_1(\alpha, \omega)=-{\rm log}_2\left(1+C_{o1}\left(\alpha, \omega\right)h^2\right)$ and $F_2(h)=e^{\frac{-h^2}{2\sigma^2}}$, $C_{i1}(\alpha, \omega)$ can be expressed as
\begin{equation}
{C_{i1}}(\alpha ,\omega ) = \int_0^{ + \infty } {{F_1}(h,\alpha ,\omega ){F_2}^\prime (h) {\rm d}h} .
\end{equation}

According to the principle of integral by parts \cite{thomas1961calculus}, $C_i(h,\alpha,\omega)$ can be further written as
\begin{equation} \label{eq:inter_c}
\begin{split}
C_{i1}\left(\alpha, \omega\right)= &{F_1}\left(h,\alpha ,\omega \right){F_2}\left(h\right)\left| {_0^{ + \infty }} \right.- \int_0^{ + \infty } {F_1}^\prime\left(h,\alpha ,\omega \right){F_2}\left(h\right){\rm{d}} h.
\end{split}
\end{equation}

Due to the fact that
\begin{equation}
	F_1(0, \alpha, \omega)F_2(0) = -{\rm log2}(1)\cdot e^0 = 0 ,
\end{equation}
\begin{equation}
\begin{split}
\mathop {\lim }\limits_{h \to  + \infty } {F_1}(h,\alpha ,\omega )F_2(h)  = \mathop {\lim }\limits_{h \to  + \infty } \left( { - {{\log }_2}\left( {1 + {C_{o1}}\left( {\alpha ,\omega } \right){h^2}} \right){e^{\frac{{ - {h^2}}}{{2{\sigma ^2}}}}}} \right)  = 0,
\end{split}
\end{equation}
(\ref{eq:inter_c}) can be expressed as
\begin{equation}
\begin{split}
&{C_{i1}}\left(\alpha ,\omega \right) 
= -\int_0^{ + \infty } {\frac{{2{C_{o1}}\left( {\alpha, \omega} \right)h}}{{\ln 2 \cdot \left( {1 + {C_{o1}}\left( {\alpha, \omega} \right){h^2}} \right)}}} {e^{ - \frac{{{h^2}}}{{2{\sigma ^2}}}}}{\rm{d}}h \\
%= &-\frac{1}{{\ln 2}}\int_0^{ + \infty } {\frac{{2h}}{{\left( {\frac{1}{{{C_{o1}}\left( {\alpha, \omega} \right)}} + {h^2}} \right)}}}  \cdot {e^{ - \frac{{{h^2}}}{{2{\sigma ^2}}}}}{\rm{d}}h \\
%{\rm{ = }}&-\frac{1}{{\ln 2}}\int_0^{ + \infty } {\frac{2\sigma^2 {e^{ - \frac{{\frac{1}{{{C_{o1}}\left( {\alpha, \omega} \right)}} + {h^2}}}{{2{\sigma ^2}}}}} {{\rm{e}}^{\frac{1}{{2\sigma^2{C_{o1}}\left( {\alpha, \omega} \right)}}}}h}{{\left( {\frac{1}{{{C_{o1}}\left( {\alpha, \omega} \right)}} + {h^2}} \right)\sigma^2}}}{\rm{d}}h \\
{\rm{ = }}&-\frac{{\rm{e}}^{\frac{1}{{2\sigma^2{C_{o1}}\left( {\alpha, \omega} \right)}}}}{\ln 2}  \int_{\frac{1}{2\sigma^2C_{o1}\left(h,\alpha,\omega)\right)}}^{ + \infty } {\frac{2\sigma^2 {e^{ - \frac{{\frac{1}{{{C_{o1}}\left( {\alpha, \omega} \right)}} + {h^2}}}{{2{\sigma ^2}}}}}}{{\left( {\frac{1}{{{C_{o1}}\left( {\alpha, \omega} \right)}} + {h^2}} \right)}}} d\left({\frac{{\frac{1}{{{C_{o1}}\left( {\alpha, \omega} \right)}} + {h^2}}}{2\sigma^2}} \right)  \\
&{\rm{ = }}-\frac{{{e^{\frac{{{N_B}}}{{4{\sigma ^2}P_s\alpha {T_N}H(\alpha ,\omega )}}}}}}{{\ln 2}}{\rm Ei}\left(- {\frac{{{N_B}}}{{4\sigma^2 P_s\alpha {T_N}H(\alpha ,\omega )}}} \right),
\end{split}
\end{equation}
where ${\rm E_i}(x)$ is the exponential integral function which is defined as
$
{\rm E_i}(x) =   \int_{-x}^{ + \infty } {\frac{{{e^{ - t}}}}{t} \rm d}t
$.

Now, by applying ${C_{o2}}\left( \alpha  \right) = -4\sigma^2 P_s\alpha {T_N}/{N_B}$, (\ref{eq:ray_aver_rate}) can be written as
\begin{equation}
{C_R}(\alpha ) =  - \frac{1}{{\pi \alpha \left( {1 + \beta } \right)\ln 2}}\underbrace {\int_0^\pi  {{e^{ - \frac{{{C_{o2}}\left( \alpha  \right)}}{{H(\alpha ,\omega )}}}}} {\rm{Ei}}\left( {\frac{{{C_{o2}}}}{{H(\alpha ,\omega )}}} \right)d\omega }_{{C_{i2}}\left( \alpha, \omega  \right)} .
\end{equation}

Then, we split $C_{i2}\left(\alpha, \omega\right)$ into several subsection integral and calculate them respectively. For $\omega \in \left[0, b_1\left(\alpha\right)\right)$, the integral can be calculated as
\begin{equation}
\begin{split}
{C_{4}}\left( {\alpha } \right) &= \int_0^{{b_1}\left( \alpha  \right)} {{e^{-\frac{{{C_{o2}}(\alpha )}}{{H(\alpha ,\omega )}}}}} {\mathop{\rm Ei}\nolimits} \left( {\frac{{{C_{o2}}}}{{H(\alpha ,\omega )}}} \right)d\omega =\int_0^{{b_1}\left( \alpha  \right)} {{e^{-\alpha C_{o2}\left(\alpha\right)}}{\rm{Ei}}\left( {{\alpha}{C_{o2}}\left( \alpha  \right)} \right)d\omega } \vspace{1em} \\
&  = \alpha \pi \left( {1 - \beta } \right){e^{-{\alpha}{C_{o2}}}}{\rm{Ei}}\left( {\alpha {C_{o2}}\left( \alpha  \right)} \right) .
\end{split}
\end{equation}

According to the \textit{Theorem} \ref{theorem_4}, for $\omega \in \left[b_1\left(\alpha\right), \pi\right]$, the integral can be written as
\begin{equation}
C_5\left( \alpha \right) =\int_{b_1\left( \alpha \right)}^{\pi}{e^{-\frac{C_{o2}(\alpha )}{H(\alpha ,\omega )}}\mathrm{Ei}\left( \frac{C_{o2}}{H(\alpha ,\omega )} \right) d\omega}\approx A_1\sum_{i=1}^N{m_i}\sqrt{1-\omega _{1i}^{2}}e^{-\frac{C_{o2}(\alpha )}{H(\alpha ,\omega _{1i})}}\mathrm{Ei}\left( \frac{C_{o2}}{H(\alpha ,\omega _{1i})} \right) .
\end{equation}
Similarly, for $\omega \in \left[b_1\left(\alpha\right), b_3\left(\alpha\right)\right)$, the integral can be written as
\begin{equation}
C_6\left( \alpha \right) =\int_{b_1(\alpha )}^{b_3(\alpha )}{e^{-\frac{C_O(\alpha )}{H(\alpha ,\omega )}}}\mathrm{Ei}\left( \frac{C_{o2}}{H(\alpha ,\omega )} \right) d\omega \approx A_2\sum_{i=1}^N{m_i\sqrt{1-\omega _{2i}^{2}}}e^{-\frac{C_{a2}(\alpha )}{H\left( \alpha ,\omega _{2i} \right)}}\mathrm{Ei}\left( \frac{C_{o2}}{H\left( \alpha ,\omega _{2i} \right)} \right)  .
\end{equation}

Finally, the SE of the proposed VPR scheme in Rayleigh channel can be written as
\begin{equation}
C_R\left( \alpha \right) =\left\{ \begin{matrix}
	\frac{1}{\alpha \pi \left( 1+\beta \right)}\left( C_4\left( \alpha \right) +C_5\left( \alpha \right) \right) ,&		b_2\left( \alpha \right) <\pi\\
	\frac{1}{\alpha \pi \left( 1+\beta \right)}\left( C_4\left( \alpha \right) +C_6\left( \alpha \right) \right) ,&		b_2\left( \alpha \right) \ge \pi\\
\end{matrix} \right. .
\end{equation}

The average SE for VPR-based secure transmission in Rayleigh channel can be obtained as
\begin{equation}  \label{eq:rayleigh_average}
C_R^\prime = \frac{1}{N_\alpha} \sum_{i=1}^{N_\alpha} C_{R^\prime}(\alpha_i).
\end{equation}
%\end{figure*}

\subsection{Theoretical SE of VPR System in Nakagami-m Channel}
Similar to (\ref{eq:ray_aver_rate}), the SE of FTN signaling in Nakagami-m channel can be formulated as
\begin{equation}
\begin{split}
{C_N}(\alpha ) = \frac{1}{{\pi \alpha \left( {1 + \beta } \right)}} \cdot  \int_0^\pi  {\underbrace {\int_0^{ + \infty } {f_N\left(h\right) \cdot {\rm{lo}}{{\rm{g}}_2}\left( {1 + \frac{{2{h^2}P_s\alpha {T_N}}}{{{N_B}}}H(\alpha ,\omega )} \right){\rm{d}}h} }_{{C_{i2}}(\alpha, \omega))}} {\rm{d}}\omega
\end{split} ,
\end{equation}
where $f_N\left(h\right)$ is the PDF of $h$ in Nakagami-m channel which can be written as  \cite{nakagami1960m}
\begin{equation}
{f_N}\left( h \right) = \frac{{2{m^m}{h^{2m - 1}}}}{{\Gamma \left( m \right){P_r}^m}}{e^{ - \frac{{m{h^2}}}{{{P_r}}}}} ,
\end{equation}
where $m$ ($m>0$) is the fading parameter , $P_r$ is the average power, $\Gamma\left(m\right)$ is the Gamma function which can be expressed as \cite{ross2004differential}
\begin{equation}
\Gamma (m) = \int_0^{ + \infty } {{t^{m - 1}}} {e^{ - t}}\;{\rm{d}}t\quad (m > 0) .
\end{equation}
 
 By applying ${C_{o3}} = {2{m^m}}/\left(\Gamma \left( m \right)P_r^m\right)$, the integral of channel gain $h$ can be written as
 \begin{equation} \label{eq:ri2}
 \begin{split}
{C_{i,2}}\left( {\alpha ,\omega } \right)  \vspace{1em} &= \int_0^{ + \infty } {{{\rm{C}}_{o3}}{{\rm{h}}^{2m - 1}}{e^{ - \frac{{m{h^2}}}{{{P_r}}}}}{\rm{lo}}{{\rm{g}}_2}\left( {1 + {C_{o1}}\left( {\alpha ,\omega } \right){h^2}} \right){\rm{d}}h} \vspace{1em} \\
& = {C_{o3}}{\cal M}\left[ {{e^{ - \frac{{m{h^2}}}{{{P_r}}}}}{\rm{lo}}{{\rm{g}}_2}\left( {1 + {C_{o1}}\left( {\alpha ,\omega } \right){h^2}} \right);2m} \right] ,
\end{split}
 \end{equation}
 where $\mathcal{M}\left[f\left(x\right);s\right]$ means the Mellin transform \cite{flajolet1995mellin} of ${f\left( x \right)}$. 
 
 \begin{theorem}  \label{theorem_5}
	(Mellin Convolution Theorem) For functions $f(x)$ and $g(x)$, it holds that
	
	\begin{equation}
	\mathcal{M} \left[ f(x)g(x);s \right] =\frac{1}{2\pi i}\!\int_{c\!-\!i\infty}^{c+i\infty}{\mathcal{M} \left[ f\left( x \right) ;s \right] \mathcal{M} \left[ g\left( x \right) ;s-u \right] du}.
	\end{equation}
\end{theorem}

Considering \textit{Theorem} \ref{theorem_5}, \eqref{eq:ri2} can be further written as
\begin{equation}  \label{eq:total}
	\begin{aligned}
	C_{i,2}\left( \alpha ,\omega \right) =\frac{1}{2\pi i}\int_{c-i\infty}^{c+\infty}{\mathcal{M} \left[ e^{-\frac{mh^2}{P_r}};2m-u \right] \mathcal{M} \left[ \log _2\left( 1+C_{o1}\left( \alpha ,\omega \right) h^2 \right) ;u \right]}du .\\
\end{aligned}
\end{equation}

\begin{property}
	The Mellin transform has the properties as follows \cite{edition2007table}.
	\begin{align}
		\mathcal{M} \left[ f\left( \nu x \right) ;s \right] =\nu ^{-s}f^*\left( s \right), \,\, \nu >0, \label{property_1} \\
		\mathcal{M} \left[ e^{-x^2};s \right] =\frac{1}{2}\Gamma \left( \frac{s}{2} \right) ,  \,\,  \mathrm{Re}\left( s \right) >0, \label{property_2} \\
		\mathcal{M} \left[ f\left( x^{\nu} \right) ;s \right] =\frac{1}{\left| \nu \right|}f^*\left( \frac{s}{\nu} \right) , \,\,  s/\nu \,\, \mathrm{is} \,\, \mathrm{feasible}, \label{property_3}  \\  
		\mathcal{M} \left[ \ln \left( 1+x \right) ;s \right] =\frac{\pi}{s\sin \left( \pi s \right)},  -1\le \mathrm{Re}\left( s \right) \le 0.  \label{property_4}
	\end{align}
\end{property}

Considering \eqref{property_1} and \eqref{property_2}, it can be obtained that
\begin{equation}  \label{eq:part1}
	\mathcal{M} \left[ e^{-\frac{mh^2}{P_r}};2m-u \right] =\left( \frac{m}{P_r} \right) ^{-\left( m-\frac{u}{2} \right)}\mathcal{M} \left[ e^{-h^2};2m-u \right] =\frac{1}{2}\left( \frac{m}{P_r} \right) ^{-\left( m-\frac{u}{2} \right)}\Gamma \left( m-\frac{u}{2} \right) .
\end{equation}

Considering \eqref{property_1}, \eqref{property_3} and \eqref{property_4}, it can be obtained that
\begin{equation}  \label{eq:part2}
\begin{aligned}
	\mathcal{M} \left[ \log _2\left( 1+C_{o1}\left( \alpha ,\omega \right) h^2 \right) ;u \right] &=\frac{1}{2\ln 2}\left[ C_{o1}\left( \alpha ,\omega \right) \right] ^{-\frac{u}{2}}\mathcal{M} \left[ \ln \left( 1+h \right) ;\frac{u}{2} \right] 
\\
&=\frac{1}{\ln 2}\left[ C_{o1}\left( \alpha ,\omega \right) \right] ^{-\frac{u}{2}}\frac{\pi}{u\sin \left( \frac{\pi u}{2} \right)}.
\end{aligned}
\end{equation}

Combining \eqref{eq:part1} and \eqref{eq:part2}, \eqref{eq:total} can be written as
\begin{equation}  \label{eq:total2}
	C_{i,2}\left( \alpha ,\omega \right) =\frac{C_{o,3}}{2\pi i}\int_{c-i\infty}^{c+i\infty}{\frac{\pi \left( \frac{P_r}{m} \right) ^{m-\frac{u}{2}}\Gamma \left( m-\frac{u}{2} \right)}{4\ln 2\cdot \left( -\frac{u}{2} \right) \left[ C_{o,1}\left( \alpha ,\omega \right) \right] ^{\frac{u}{2}}\sin \left( -\frac{\pi u}{2} \right)}}du.
\end{equation}

\begin{property}  \label{property_gamma}
	Gamma function has the properties as follows.
	\begin{align}	
		\Gamma \left( 1-x \right) \Gamma \left( x \right) =\frac{\pi}{\sin \left( \pi x \right)},  \label{property_5}  \\
		\Gamma \left( x+1 \right) =x\Gamma \left( x \right)  .\label{property_6}
	\end{align}
\end{property}

Considering \textit{Property} \ref{property_gamma}, \eqref{eq:total2} can be further written as
\begin{equation}
	\begin{aligned}
	C_{i,2}\left( \alpha ,\omega \right) =&\frac{C_{o,3}}{2\pi i}\int_{c-i\infty}^{c+i\infty}{\frac{\left( \frac{P_r}{m} \right) ^{m-\frac{u}{2}}\Gamma \left( m-\frac{u}{2} \right) \Gamma \left( 1+\frac{u}{2} \right) \Gamma \left( -\frac{u}{2} \right) \Gamma \left( -\frac{u}{2} \right)}{4\ln 2\cdot \Gamma \left( 1-\frac{u}{2} \right) \cdot \left[ C_{o,1}\left( \alpha ,\omega \right) \right] ^{\frac{u}{2}}}}du\\
	=&\frac{C_{o,3}\left( \frac{P_r}{m} \right) ^m}{2\ln 2}\cdot \frac{1}{2\pi i}\int_{c-i\infty}^{c+i\infty}{\frac{\Gamma \left( m-\frac{u}{2} \right) \Gamma \left( 1+\frac{u}{2} \right) \Gamma \left( -\frac{u}{2} \right) \Gamma \left( -\frac{u}{2} \right)}{\Gamma \left( 1-\frac{u}{2} \right)}}\left( \frac{m}{P_rC_{o1}\left( \alpha ,\omega \right)} \right) ^{\frac{u}{2}}d\left( \frac{u}{2} \right) \\
	=& \frac{C_{o,3}\left( \frac{P_r}{m} \right) ^m}{2\ln 2}\cdot G_{1,0}^{3,1}\left( \left. \begin{array}{c}
	0,1\\
	0,0,m\\
	\end{array} \right|\frac{m}{P_rC_{o1}\left( \alpha ,\omega \right)} \right),
\end{aligned}
\end{equation}
 where $G_{p,\;q}^{m,\;n}\left( {\begin{array}{*{20}{c}}
 	{{a_1},{a_2} \cdots {a_p}}\\
 	{{b_1},{b_2} \cdots {b_q}}
 	\end{array}\left| z \right.} \right)$ represents the Meijer-G function \cite{bateman1953higher}.
 
 By applying $C_{o4}={m{N_B}}/\left({2P_s\alpha {T_N}{P_r}}\right)$, for $\omega \in\left[0, b_{1}(\alpha)\right)$, the integral can be written as
 \begin{equation}
 \begin{split}
 {C_{7}}(\alpha ) &= \int_0^{{b_1}(\alpha )} {\frac{1}{{\Gamma \left( m \right)\ln (2)}} \cdot G_{1,\;0}^{3,\;1}\left( {\left. {\begin{array}{*{20}{c}}
 				{0,1}\\
 				{0,0,m}
 				\end{array}} \right|\alpha {C_{o4}}\left( {\alpha ,\omega } \right)} \right)} {\rm{d}}\omega \\
    & = \frac{\alpha \pi \left(1-\beta\right)}{{\Gamma \left( m \right)\ln (2)}} \cdot G_{1,\;0}^{3,\;1}\left( {\left. {\begin{array}{*{20}{c}}
    		{0,1}\\
    		{0,0,m}
    		\end{array}} \right|\alpha {C_{o4}}\left( {\alpha ,\omega } \right)} \right).
 	\end{split}
 \end{equation}
 
 For $\omega \in \left[b_{1}(\alpha), \pi\right)$, the integral can be written as
 \begin{equation}
 \begin{aligned}
	C_8(\alpha )&=\int_{b_1(\alpha )}^{\pi}{\frac{1}{\Gamma \left( m \right) \ln\mathrm{(}2)}\cdot G_{1,\;0}^{3,\;1}\left( \begin{array}{c}
	0,1\\
	0,0,m\\
\end{array}\left| \frac{C_{o4}}{H\left( \alpha ,\omega \right)} \right. \right)}d\omega\\
	&\approx \frac{C_1}{\Gamma \left( m \right) \ln\mathrm{(}2)}\sum_{i=1}^N{m_i}\sqrt{1-\omega _{1i}^{2}}\cdot G_{1,\;0}^{3,\;1}\left( \begin{array}{c}
	0,1\\
	0,0,m\\
\end{array}\left| \frac{C_{o4}}{H\left( \alpha ,\omega _{1i} \right)} \right. \right) .\\
\end{aligned}
 \end{equation}
 
 And for $\omega \in\left[b_{1}(\alpha), b_{3}(\alpha)\right)$, the integral can be written as
 \begin{equation}
 \begin{aligned}
	C_9(\alpha )&=\int_{b_1(\alpha )}^{b_3\left( \alpha \right)}{\frac{1}{\Gamma \left( m \right) \ln\mathrm{(}2)}\cdot G_{1,\;0}^{3,\;1}\left( \begin{array}{c}
	0,1\\
	0,0,m\\
\end{array}\left| \frac{C_{o4}}{H\left( \alpha ,\omega \right)} \right. \right)}d\omega\\
	&\approx \frac{C_2}{\Gamma \left( m \right) \ln\mathrm{(}2)}\sum_{i=1}^N{m_i}\sqrt{1-\omega _{2i}^{2}}\cdot G_{1,\;0}^{3,\;1}\left( \begin{array}{c}
	0,1\\
	0,0,m\\
\end{array}\left| \frac{C_{o4}}{H\left( \alpha ,\omega _{2i} \right)} \right. \right) .\\
\end{aligned}
 \end{equation}
 
 Finally, the SE of the proposed VPR system in Nakagami-m channel can be written as
 \begin{equation}
 C_N\left( \alpha \right) =\begin{cases}
	\frac{1}{\pi \alpha \left( 1+\beta \right)}\left( C_7(\alpha )+C_8(\alpha ) \right) ,&		b_2(\alpha )<\pi\\
	\frac{1}{\pi \alpha \left( 1+\beta \right)}\left( C_7(\alpha )+C_9(\alpha ) \right) ,&		b_2(\alpha )\ge \pi\\
\end{cases}.
 \end{equation}

 The average SE of the VPR-based secure system in Nakagami-m channel can be written as
 \begin{equation} \label{eq:nakagami_average}
 	C_N^\prime = \frac{1}{N_\alpha} \sum_{i=1}^{N_\alpha} C_N(\alpha_i).
 \end{equation}

\section{Numerical Results} \label{sec:numerical}
This section carries out comprehensive analysis and evaluation for the proposed VPR transmission systems. The simulation employs the binary phase shift keying (BPSK) modulation and SRRC filter with roll-off factor $\beta$. And the training parameters for the DNN in the proposed simplified symbol packing ratio estimation are listed in Table \ref{tab:param}. Each group mentioned in the table consists of 20 received symbols.

\begin{table}[ht]
	\centering
	\renewcommand\arraystretch{1}
	\caption{Training and testing parameters of the DNN in the proposed simplified symbol packing ratio estimation.}
	\label{tab:param}
\begin{tabular}{|c|c|c|c|}
	\hline 
	item & value & item & value \\ 
	\hline 
	number of neurons & (20, 1000, 500, 250, 1) & loss function & mean square error (MSE) \\ 
	\hline 
	training data size & $3\times 10^6$ groups & learning rate & 0.001 \\ 
	\hline 
	training $E_b/N_0$ & 4dB & start / end sparsity & 0 / 0.5 \\
	\hline
	training epoch & 50 & testing data size & $3\times 10^6$ groups \\ 
	\hline 
	optimizer & Adam & & \\ 
	\hline
\end{tabular}
\end{table}

\subsection{SE of the Proposed VPR System in AWGN Channel}
\begin{figure}[ht]
\centering
\subfloat[$\beta=0.5$, $\alpha_{Th}\approx0.667$.]{
	\includegraphics[width=0.3\linewidth]{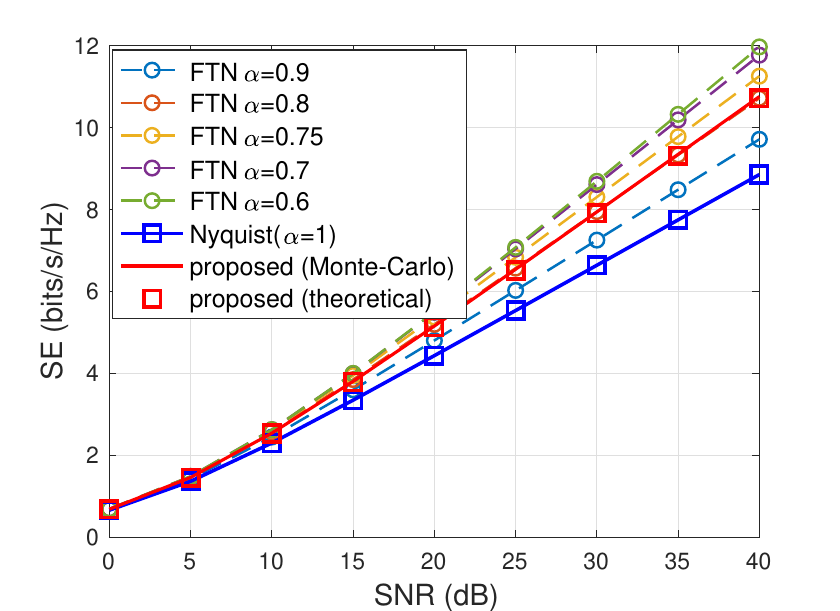}
	\label{fig:rate_0p5}}
\hspace{0.8in}
\subfloat[$\beta=0.3$, $\alpha_{TH}\approx0.769$.]{
	\includegraphics[width=0.3\linewidth]{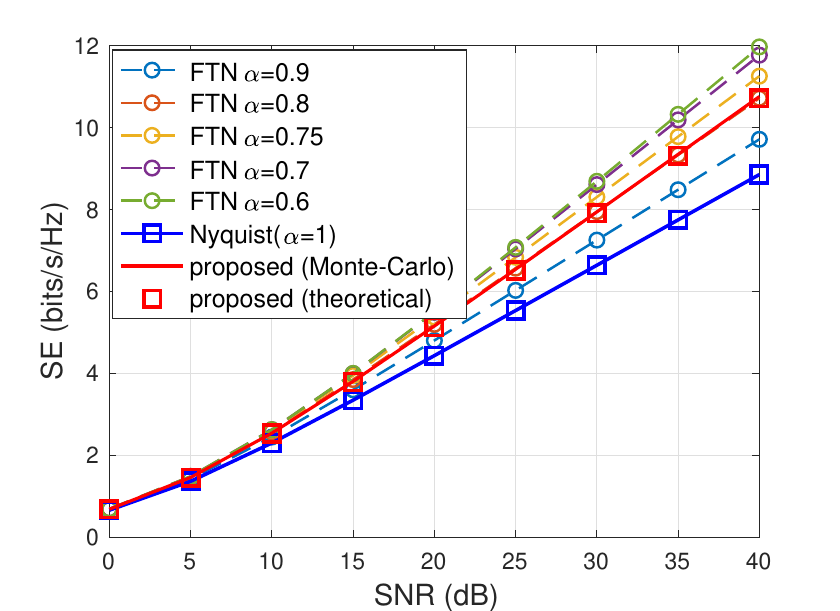}
	\label{fig:rate_0p3}}
\quad
\caption{SE of the proposed VPR system versus Nyquist-criterion transmission in AWGN channels.}
\label{fig:rate_awgn}
\end{figure}

The average SEs of the proposed VPR system in AWGN channel are illustrated in Fig. \ref{fig:rate_awgn}\subref{fig:rate_0p5} and Fig. \ref{fig:rate_awgn}\subref{fig:rate_0p3} with roll-off factors $\beta=0.5$ and $\beta=0.3$ respectively. The curves labeled \textit{Monte-Carlo} or without special label are obtained by numerical simulation. While the curve labeled \textit{theoretical} is calculated by (\ref{eq:awgn_average}). To avoid the confusion resulting from too many curves and marks, only the curve for average theoretical capacity in Section \ref{sec:rate} is plotted. And the perfect match of the results by theoretical derivation and numerical simulation proves the correctness of the SE presented in Section \ref{sec:rate}.

It should be noticed that, the SE of FTN signaling only increases when $\alpha>1/(1+\beta)$, which has been proved by \cite{rusek2009constrained}. So, the curves with $\alpha \le 1/(1+\beta)$ coincide and show the same SE, as demonstrated in the figures. To make it more clearly, we add the threshold $\alpha_{Th}=1/(1+\beta)$ in the subtitles of each figure.

% Also, as can be seen from the figures, by employing FTN signaling with different $\alpha$ values, the proposed system can achieve up to 20\% higher average spectrum efficiency gain beyond conventional Nyquist-criterion systems, without any extra spectrum resources required. It can obviously improve the utilization of the precious spectrum resources and meet the increasing demand from communication businesses on the data traffic. 

%Also, as can be seen from the comparison between Fig. \ref{fig:rate_0p5} and Fig. \ref{fig:rate_0p3}, the SE gain will decrease when a smaller roll-off factor $\beta$ is employed.
%As can be seen, in the FTN signaling, the transmission rate does not always increase with the increase of $\alpha$ value. Such a phenomenon is mainly resulting from the combined effects of both smaller symbol duration and severe ISI. For example, when $\alpha=0.7$, although the symbol duration is smaller than that in other scenarios, the severe ISI produces a more obvious performance degradation on the final transmission rate. This is also what \emph{Mazo limit} studies on.

% 原来是是0.82
% 这段话可以复制到后面的部分，暂时注释
% Although the analysis and the simulation have proved the spectrum efficiency gain of the proposed scheme, an effective blind estimation for packing ratio is required to make the communications available. Fig. \ref{fig:awgn_4dB} illustrates the accuracy performance of the proposed packing ratio estimation in AWGN under SNR = 4dB. $\alpha$ is the correct packing ratio of the input data. Every grid represents the probability of outputting $1$ when the estimation for whether $\alpha = \alpha_k$ is implemented. It should be noticed that the estimations for all $\alpha$ values are carried out independently and the $\alpha_k$ with the most 1 output is considered the correct packing ratio of the data. Hence, the sum value of any row or column in Fig. \ref{fig:awgn_4dB} does not have to be 1.

%As can be seen from the figure, for the data with any actual packing ratio value $\alpha$, the estimation for $\alpha_k=\alpha$ always outputs the most 1 values so that the system can choose the correct $\alpha_k$ as the estimated packing ratio value. This has proved the effectiveness of the proposed simplified estimation algorithm.

\subsection{SE of the Proposed VPR System in Rayleigh and Nakagami-m Channels}
\begin{figure}[ht]
\centering
\subfloat[Rayleigh channel, $\beta=0.5$, $\alpha_{Th}\approx0.667$.]{
	\includegraphics[width=0.3\linewidth]{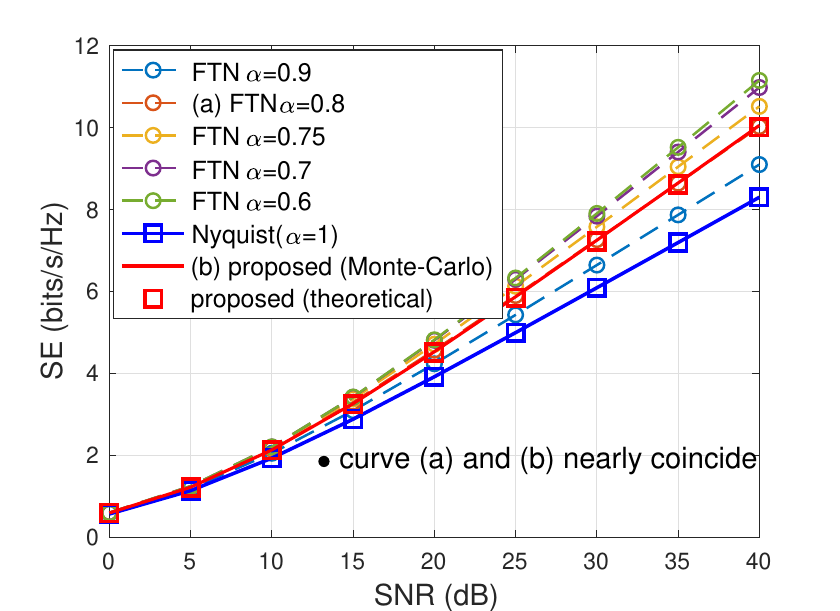}
	\label{fig:rayleigh_rate}}
\hspace{0.8in}
\subfloat[Nakagami-m channel, $\beta=0.5$, $m=3$, $P_r=2$, and $\alpha_{Th}\approx0.667$.]{
	\includegraphics[width=0.3\linewidth]{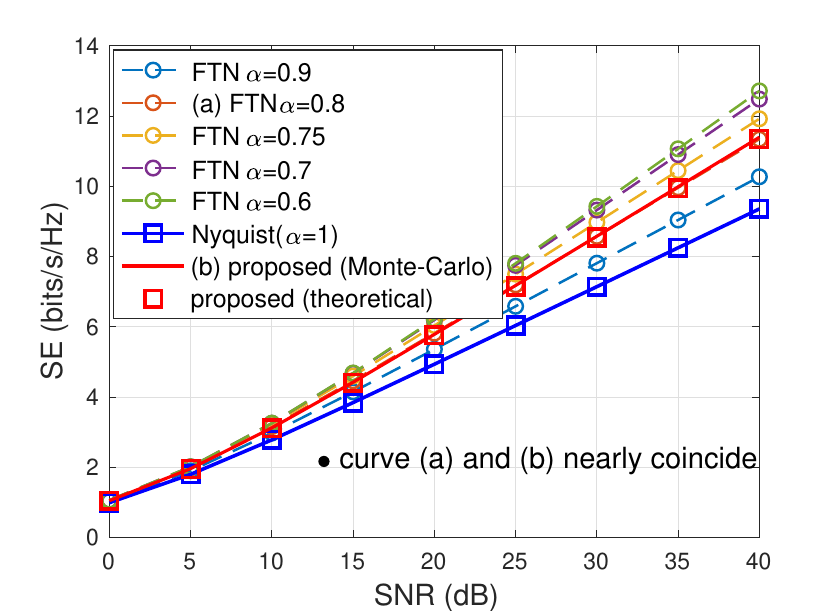}
	\label{fig:nakagami_rate}}
\caption{SE of the proposed VPR system versus Nyquist-criterion transmission in Rayleigh and Nakagami-m channels.}
\label{fig:rate_other}
\end{figure}

Fig. \ref{fig:rate_other}\subref{fig:rayleigh_rate} and Fig. \ref{fig:rate_other}\subref{fig:nakagami_rate} illustrate the SE of the proposed scheme in Rayleigh and Nakagami-m ($m=3$ and $P_r=2$) channels. The curve labeled \textit{Monte-Carlo} is obtained by independent repeated trials with randomly generated channel gain $h$ values. And the curve labeled \textit{theoretical} is calculated by (\ref{eq:rayleigh_average}) and (\ref{eq:nakagami_average}). 7 points are considered for the Chebyshev-Gauss quadrature.

As can be seen, the Monte-Carlo simulation fits the curve with theoretical result provided in Section \ref{sec:rate}. It shows that (\ref{eq:rayleigh_average}) and (\ref{eq:nakagami_average}) accurately describe the capacities of VPR scheme in Rayleigh and Nakagami-m channels.

\subsection{Performance of the Proposed Simplified Estimation for FTN Signaling in Different Channels}

%\begin{figure*}[ht!]
%	\centering
%	\includegraphics[width=0.8\linewidth]{heatmap.pdf}
%	\caption{The accuracy of the proposed simplified estimation for packing ratio of FTN signaling. (a) AWGN channel, Eb/No=4dB; (b) Rayleigh channel, Eb/No=25dB; (c) Nakagami-m channel, Eb/No=18dB. }
%	\label{fig:heatmap}
%\end{figure*}

\begin{figure*}[ht!]
	\centering
	\subfloat[AWGN, Eb/No=4dB.]{
		\includegraphics[width=0.26\textwidth]{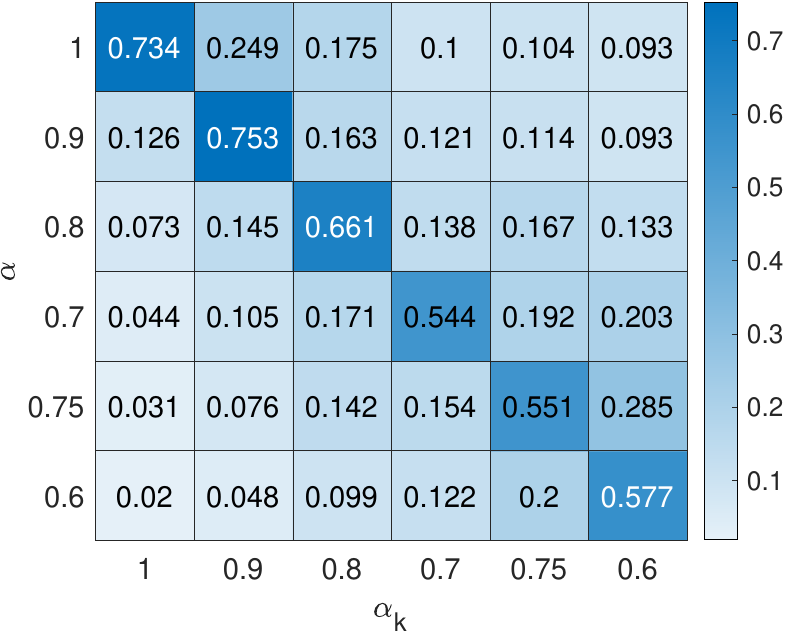}
		\label{fig:ber_0p5_c1}}
	\subfloat[Rayleigh, Eb/No=25dB.]{
		\includegraphics[width=0.26\textwidth]{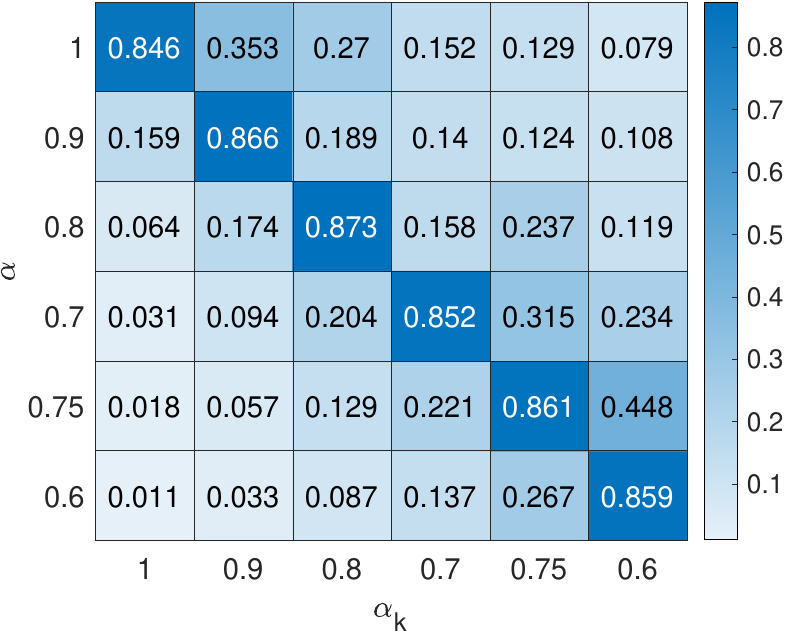}
		\label{fig:ber_0p5_c1}}
	\subfloat[Nakagami-m, Eb/No=18dB.]{
		\includegraphics[width=0.26\textwidth]{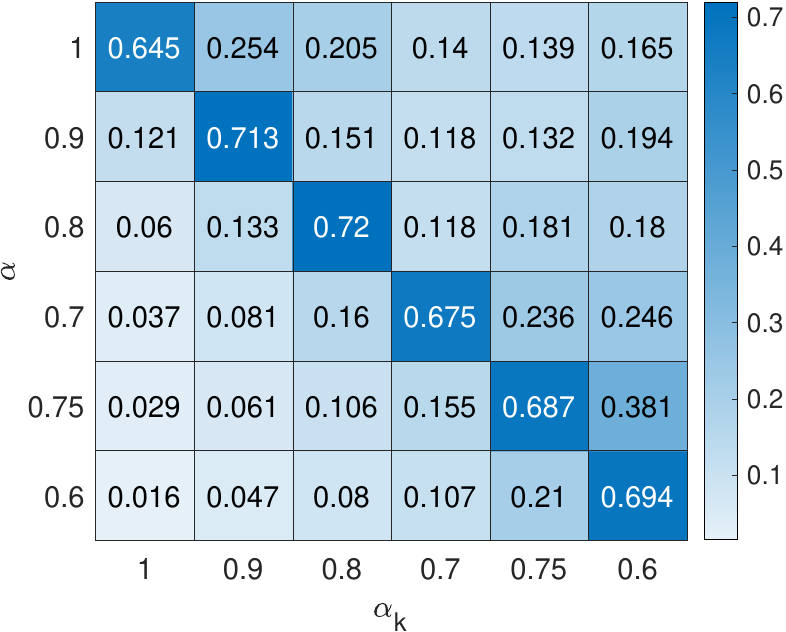}
		\label{fig:ber_0p5_c1}}
	\caption{The accuracy of the proposed simplified estimation for packing ratio of FTN signaling.}
	\label{fig:heatmap}
\end{figure*}

For the proposed scheme, an effective blind estimation for the packing ratio is required to make the communications available. Fig. \ref{fig:heatmap} illustrates the accuracy of the proposed packing ratio estimation in different channels. $\alpha$ is the real packing ratio of the input data. Every grid represents the probability of outputting $1$ in the estimation branch for whether $\alpha = \alpha_k$. It should be noticed that the estimations for all $\alpha$ values are carried out independently and the $\alpha_k$ with the most $1$ output is considered the correct packing ratio of the data. Hence, the sum value of any row or column in Fig. \ref{fig:heatmap} does not have to be 1.

As seen, the correct $\alpha$ value always corresponds to the highest probability to output 1. After a specific time to count the number of 1 in each branch, the right $\alpha_k$ will be chosen as the estimated $\alpha$ value. Hence, the simplified estimation for $\alpha$ is proved to be effective.

\subsection{SE Gain of the Proposed VPR-based Scheme over Conventional Nyquist Transmissions}

%todo 这个地方需要重新绘图

\begin{table*}[ht!]
	\caption{The simulated packing ratio for different channels and SNRs}
	\label{tab:snrrange}
	\centering
	\renewcommand\arraystretch{1}
	\setlength{\tabcolsep}{4mm}
	\begin{tabular}{c|c|c|c|c|c|c}
		\hline
		Eb/No range (dB) & $\alpha=1.0$ & $\alpha=0.9$ & $\alpha=0.8$ & $\alpha=0.75$ & $\alpha=0.7$ & $\alpha=0.6$ \\
		\hline
		AWGN ($\beta=0.5$) & -- & -- & $(-\infty, 7]$ & $(7, 7.1]$ & $(7.1, 8.6]$ & $(8.6, +\infty)$ \\
		\hline
		AWGN ($\beta=0.4$) & -- & --& $(-\infty, 7.1]$ & $(7.1, 7.3]$ & $(7.3, 9.9]$ & $(9.9, +\infty)$ \\
		\hline
		AWGN ($\beta=0.3$) & -- & --& $(-\infty, 7.6]$ & $(7.6, 8.6]$ & $(8.6, 12.1]$ & $(12.1, +\infty)$ \\
		\hline
		Rayleigh ($\beta=0.5$) & -- & $(-\infty, 23.5]$ & $(23.5, 24.1]$ & $(24.1, 24.4]$ & $(24.4, 25.1]$ & $(25.1, +\infty)$ \\
		\hline
		Rayleigh ($\beta=0.4$) & -- & $(-\infty, 24.6]$ & $(24.6, 25.3]$ & $(25.3, 25.6]$ & $(25.6, 26.9]$ & $(26.8, +\infty)$ \\
		\hline
		Rayleigh ($\beta=0.3$) & -- & $(-\infty, 25.2]$ & -- & $(25.2, 26.2]$ & $(26.2, 27]$ & $(27, +\infty)$ \\
		\hline
		Nakagami-m ($\beta=0.5$) & -- & $(-\infty, 11.2]$ & $(11.2, 12]$ & $(12, 12.6]$ & $(12.6, 13.2]$ & $(13.2, +\infty)$ \\
		\hline
		Nakagami-m ($\beta=0.4$) & -- & --  & $(-\infty, 12]$ & $(12, 12.6]$ & $(12.6, 14]$ & $(14, +\infty)$ \\
		\hline
		Nakagami-m ($\beta=0.3$) & -- & -- & $(-\infty, 12]$ & $(12, 13]$ & $(13, 15]$ & $(15, +\infty)$ \\
		\hline 
	\end{tabular}
\end{table*}

%\begin{figure*}[ht!]
%	\centering
%	\includegraphics[width=0.9\linewidth]{gain1.pdf}
%	\label{fig:gain1}
%\end{figure*}

%\begin{figure*}[ht!]
%	\centering
%	\includegraphics[width=0.7\linewidth]{total_gain.pdf}
%	\caption{Comparison for SE of the proposed scheme and Nyquist scheme. (a) AWGN, $\beta=0.5$; (b) AWGN, $\beta=0.4$; (c) AWGN, $\beta=0.3$; (d) Rayleigh, $\beta=0.5$; (e) Rayleigh, $\beta=0.4$; (f) Rayleigh, $\beta=0.3$; (h) Nakagami, $\beta=0.5$; (i) Nakagami, $\beta=0.4$; (j) Nakagami, $\beta=0.3$}
%	\label{fig:gain2}
%\end{figure*}

\begin{figure*}[ht!]
	\centering
	\subfloat[AWGN, $\beta=0.5$.]{
		\includegraphics[width=0.16\textwidth]{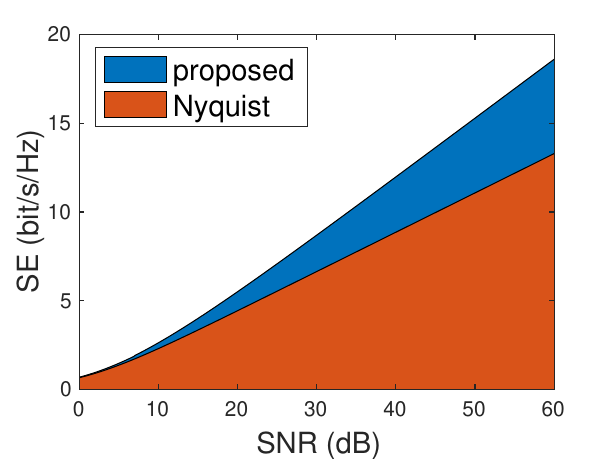}
		\label{fig:ber_0p5_c1}}
	\hspace{0.07in}
	\subfloat[AWGN, $\beta=0.4$.]{
		\includegraphics[width=0.16\textwidth]{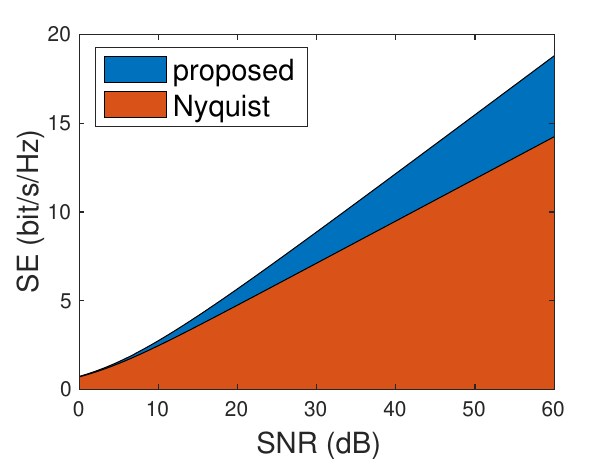}
		\label{fig:ber_0p5_c1}}
	\hspace{0.07in}
	\subfloat[AWGN, $\beta=0.3$.]{
		\includegraphics[width=0.16\textwidth]{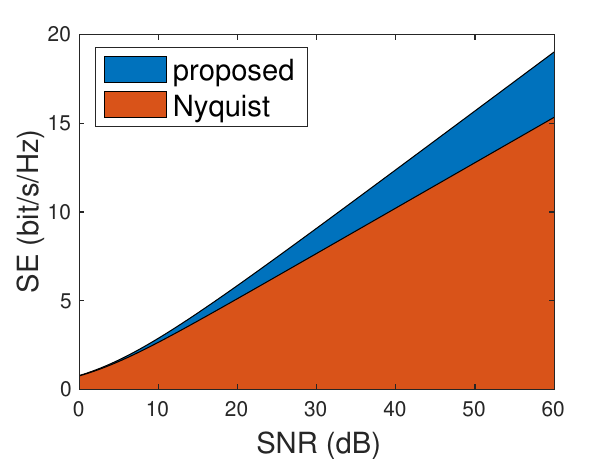}
		\label{fig:ber_0p5_c1}}
	\hspace{0.07in}
	\subfloat[Rayleigh, $\beta=0.5$.]{
		\includegraphics[width=0.16\textwidth]{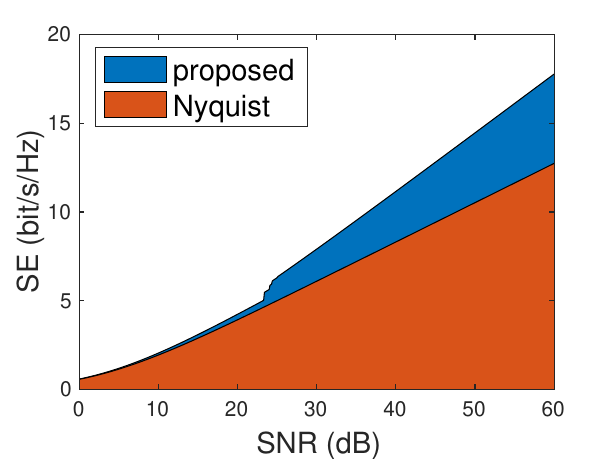}
		\label{fig:ber_0p5_c1}}
	\hspace{0.07in}
	\subfloat[Rayleigh, $\beta=0.4$.]{
		\includegraphics[width=0.16\textwidth]{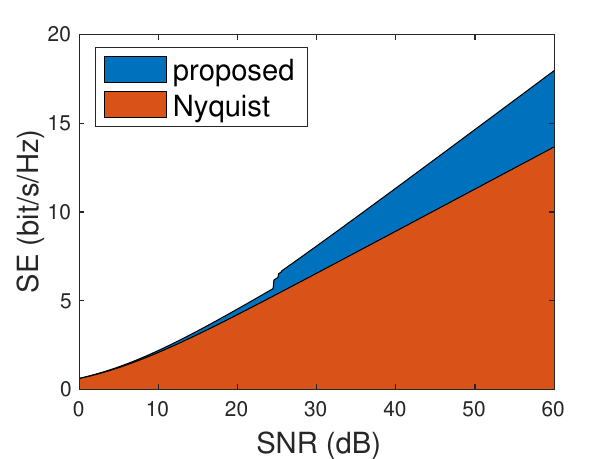}
		\label{fig:ber_0p5_c1}}
	\\
	\subfloat[Rayleigh, $\beta=0.3$.]{
		\includegraphics[width=0.17\textwidth]{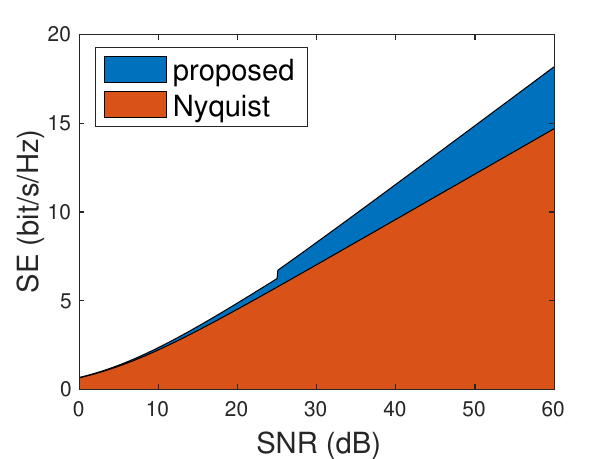}
		\label{fig:ber_0p5_c1}}
	\hspace{0.15in}
	\subfloat[Nakagami,$\beta=0.5$.]{
		\includegraphics[width=0.17\textwidth]{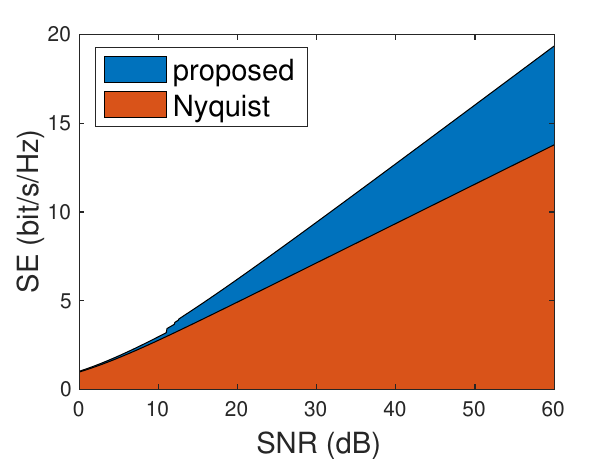}
		\label{fig:ber_0p5_c1}}
	\hspace{0.15in}
	\subfloat[Nakagami,$\beta=0.4$.]{
		\includegraphics[width=0.17\textwidth]{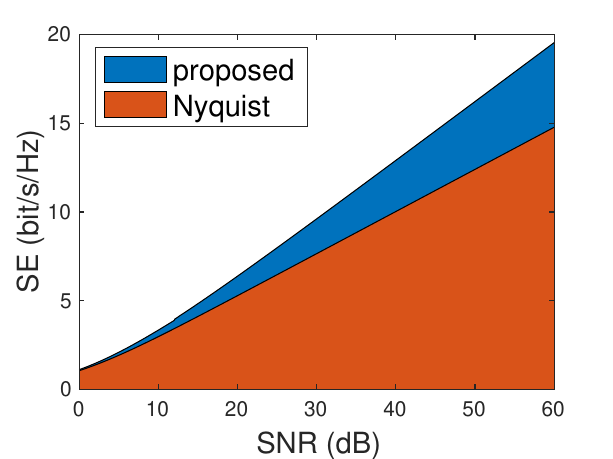}
		\label{fig:ber_0p5_c1}}
	\hspace{0.15in}
	\subfloat[Nakagami,$\beta=0.3$.]{
		\includegraphics[width=0.17\textwidth]{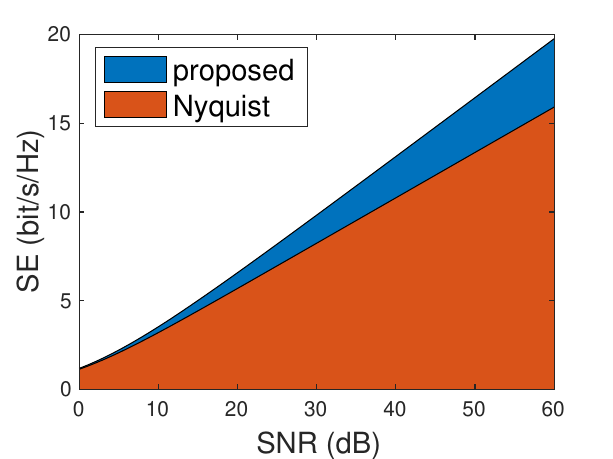}
		\label{fig:ber_0p5_c1}}
	\caption{Comparison for SE of the proposed scheme and Nyquist scheme.}
	\label{fig:gain2}
\end{figure*}

In this part, we provide an example of implementation for the proposed VPR-based high SE transmission, where the maximum a priori probability (MAP) \cite{li2018reduced} is employed as the detection algorithm, as shown in Table \ref{tab:snrrange}. The parameters for the Nakagami-m channel here are set as $\mu=3$ and $\omega=2$. Under a certain SNR,  we will choose the smallest one of the optional $\alpha$ values with which the BER is lower than $10^{-3}$ to achieve the highest SE. And to better compare the SE gain in different channels, the simulated SNR range is set as $\left[0, 60\right]$ (dB) for all scenarios.

Fig. \ref{fig:gain2} detailed illustrates the SE comparison between the proposed scheme and the conventional Nyquist system. Obvious SE gain, as seen, can be achieved by the proposed VPR system under all simulated channels and roll-off factors. A flexible switching strategy can help the system take advantage of high SNR to achieve a higher SE up to 47\% without any extra spectrum consumed. The application of the proposed scheme in NOMA and multi-beam satellite, as mentioned before, can be designed with the similar simulations or experiments.

\subsection{BER performance for Alice-Bob and Alice-Eve}

\begin{figure}[ht]
	\centering
	\includegraphics[width=0.4\linewidth]{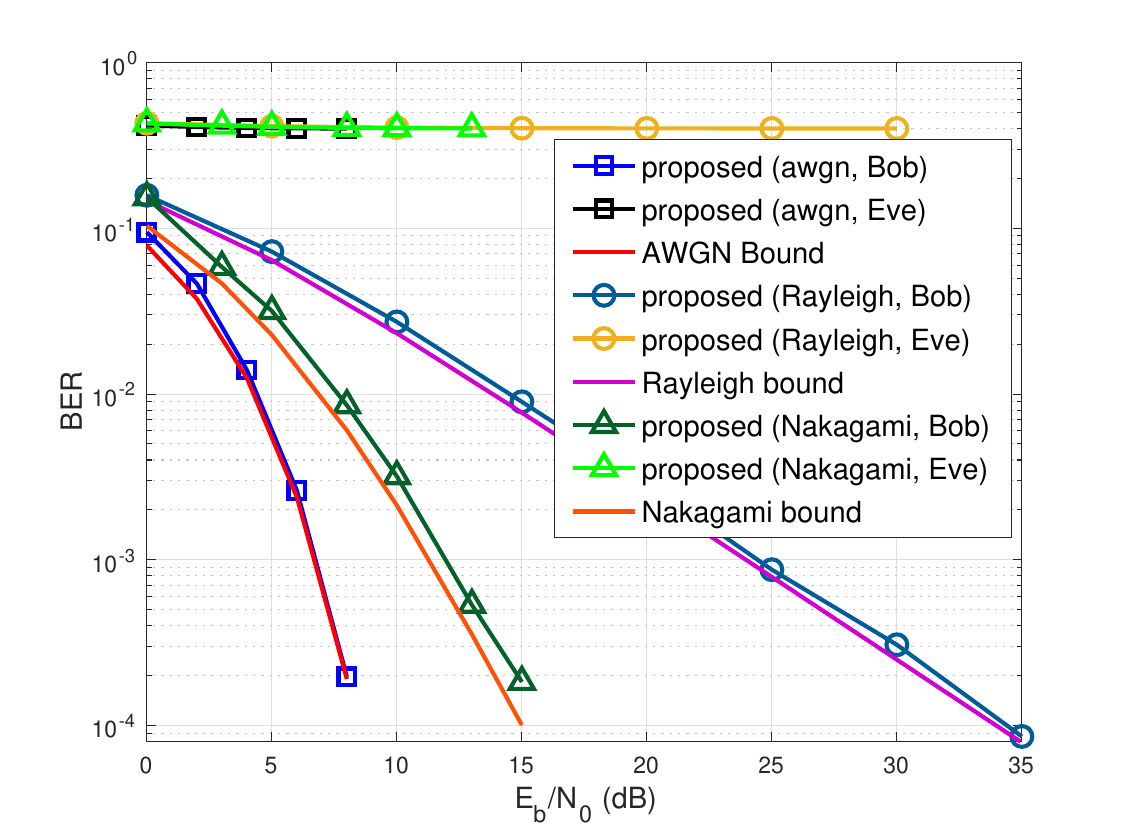}
	\caption{BER performance of the proposed system for Alice-Bob and Alice-Eve links.}
	\label{fig:ber}
\end{figure}

Fig \ref{fig:ber} demonstrates the BER performance of Alice-Bob and Alice-Eve links. As can be seen, the Alice-Bob link can achieve nearly the same BER performance as that in the ISI-free AWGN channel. For the Alice-Eve link, when $\alpha_E\ne\alpha_A$, it will not be able to sample the received signals by the expected interval. Despite the assumption that when $\alpha_E=\alpha_A$, sampling offset is not taken into consideration, the average BER of the Alice-Eve link is still poor enough.

\subsection{The Power of Random Segment Starting Positions to Avoid Attack and Detection}

\begin{figure}[!ht]
	\begin{center}
	\includegraphics[width=0.3\textwidth]{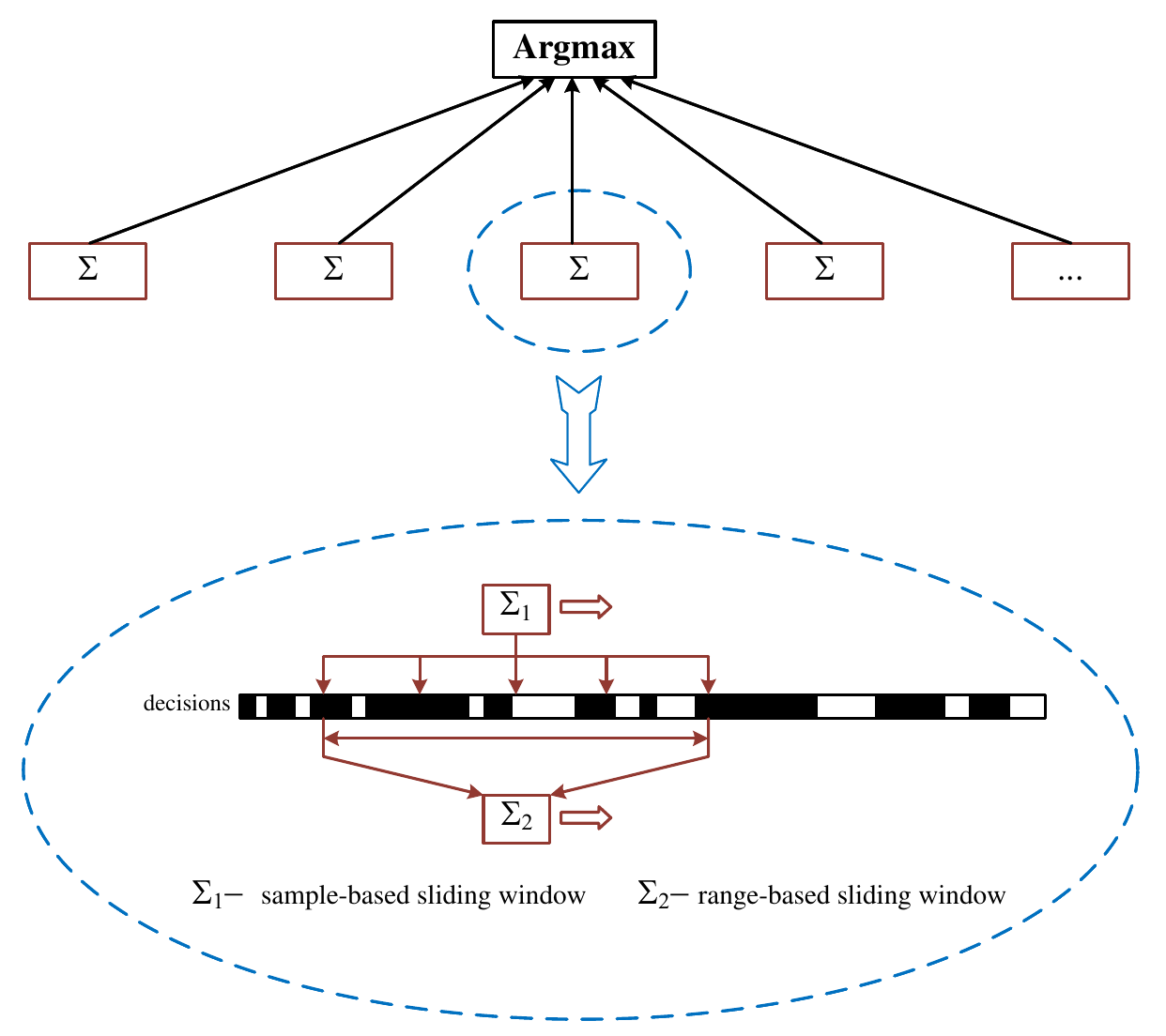}
	\end{center}
	\caption{The sample-based and range-based sliding windows for the simulation of the proposed estimation.}
	\label{fig:sum_scheme_c1}
\end{figure}

\begin{figure}[ht]
	\centering
	\subfloat[Sample-based sliding window.]{
		\includegraphics[width=0.46\textwidth]{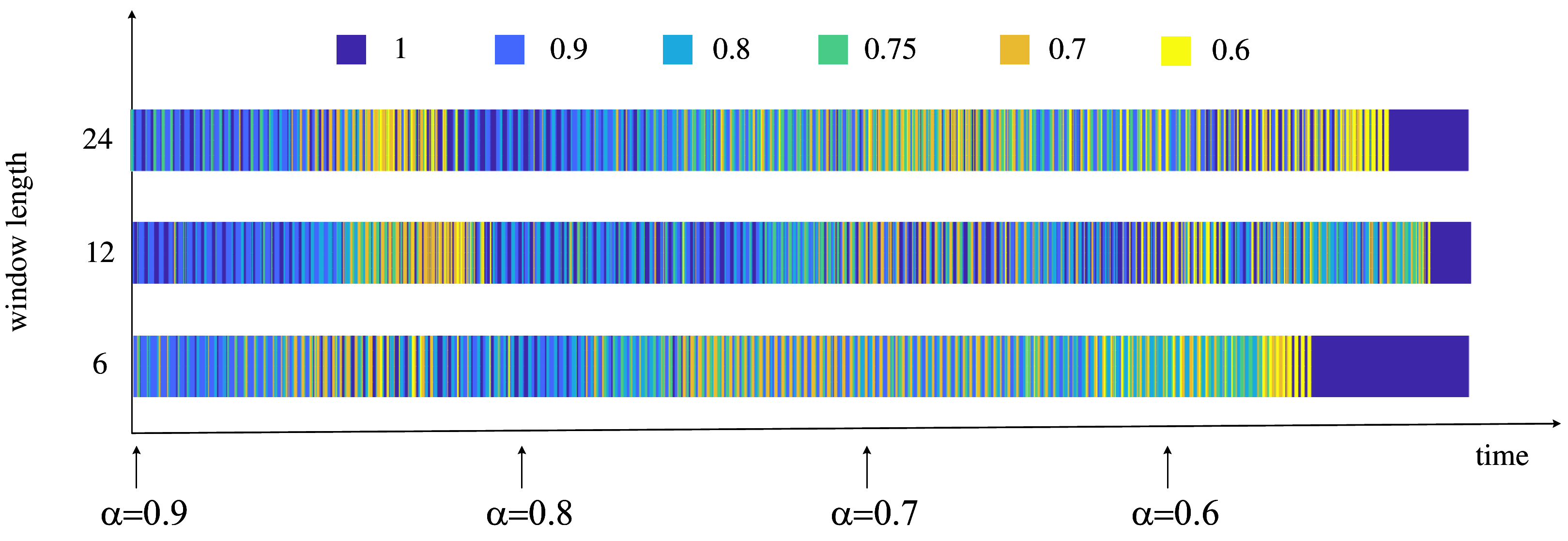}
		\label{fig:result_new_c1}}
		\hspace{0.5em}
	\subfloat[Range-based sliding window.]{
		\includegraphics[width=0.46\textwidth]{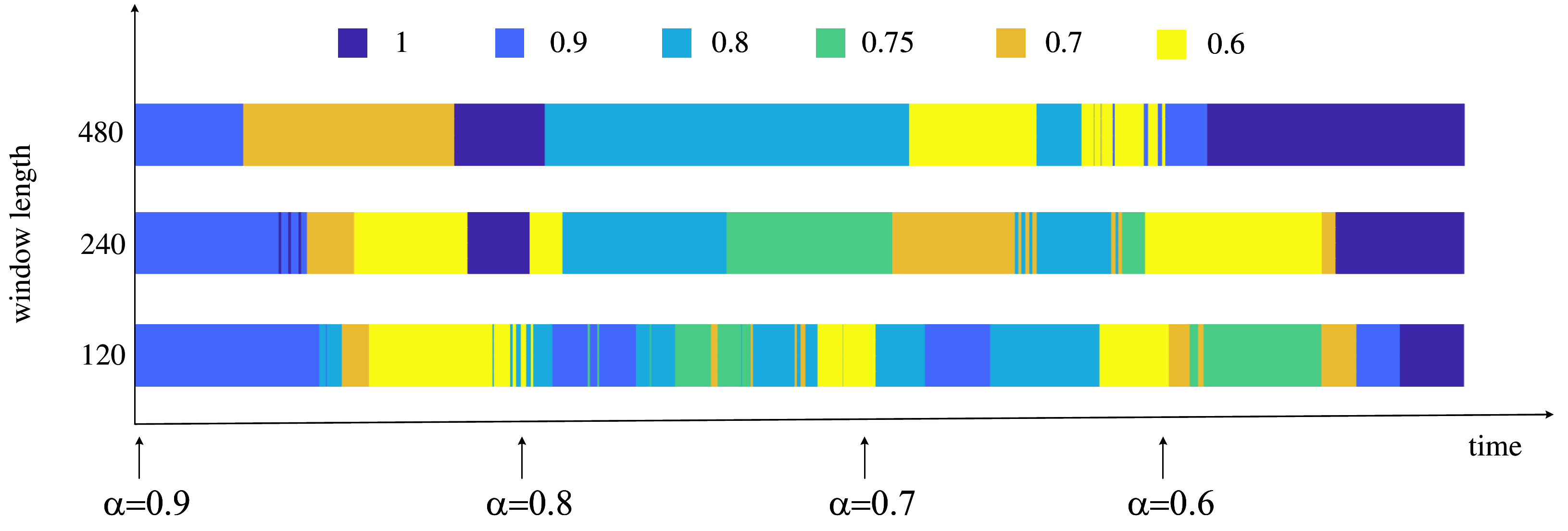}
		\label{fig:result_old_c1}}
	\caption{Estimation by Eve at the exampled frame.}
	\label{fig:result_est}
\end{figure}

Eve's estimations on the exampled frame with sample-based and range-based sliding windows (presented in Fig. \ref{fig:sum_scheme_c1}) are demonstrated in Fig. \ref{fig:result_est}\subref{fig:result_new_c1} and Fig. \ref{fig:result_est}\subref{fig:result_old_c1}, respectively. An up-sampling with 20 times is employed. And the frame is constructed with $\alpha=0.9, 0.8, 0.7, 0.6$, where $\alpha$ for each segment has been marked in the figures. 

%\begin{figure}[!ht]
%	\begin{center}
%		\includegraphics[width=0.5\textwidth]{security_result_new.pdf}
%	\end{center}
%	\caption{Estimation by Eve at the exampled frame with sample-based sliding window.}
%	\label{fig:result_new_c1}
%\end{figure}
%
%\begin{figure}[!ht]
%	\begin{center}
%		\includegraphics[width=0.5\textwidth]{security_result_old.pdf}
%	\end{center}
%	\caption{Estimation by Eve at the exampled frame with range-based sliding window.}
%	\label{fig:result_old_c1}
%\end{figure}

The estimation in Fig. \ref{fig:result_est}\subref{fig:result_new_c1} is based on $\eta_1$ continuous decisions with fixed interval $\alpha T_N$, where $\eta_1$ is the length of the sliding window. As shown, the result is messy and it is difficult to find a pattern to map the estimation to the original packing ratio for each segment. In fact, the information of starting positions  helps the receiver carry out the estimation at the perfect times to eliminate the interference of other erroneous results.
	
Another way for Eve's estimation is to employ the continuous $\eta_2$ decisions with fixed interval $\alpha T_N/U_s$, where $U_s$ is the up-sampling times. As shown in Fig. \ref{fig:result_est}\subref{fig:result_old_c1}, the estimation is still confusing. And especially, the starting position cannot be inferred by the estimation results.

\subsection{BER Degradation of VPR-based Secure Transmission}

	According to the simulation results, the VPR-based secure transmission performs nearly the same SE with FTN signaling where $\alpha=0.8$, $\beta=0.5$ and $\alpha=0.9$, $\beta=0.3$. So, we compare the BER of them  under such two cases, where the following channel codings are considered. 
	
	\begin{itemize}
	\item \textbf{Low density parity check (LDPC) code.}
	We employ the (1296, 648) LDPC code with a rate of 1/2. The parity matrix is defined in \cite{9502043}. And the back propagation (BP) is employed as the decoding algorithm.
	
	\item \textbf{Turbo code.}	
	We employ the (6298, 1256) Turbo code with a rate of 628/3149 and the constraint length of 4. The parity bits are obtained by $y_1=x^3+x^1+x^0$ and $y_2=x^3+x^2+x^1+x^0$, where $x^{\kappa}$ represents the $\kappa$-th bits in the state of shift registers. And the feedback bit is calculated by $r_b=x^3+x^2+x^0$. The MAP is applied as the decoding algorithm.

	\item \textbf{Convolutional code (CC).}
	We employ the (3768, 1256) CC with a rate of 1/3. The constraint length and the structure of the shift registers are the same as that of the turbo code presented in the previous item. No tail bits are required in this case. And Viterbi decoding with hard decisions is employed as the decoding algorithm.
	\end{itemize}
	
\begin{figure}[!ht]
	\subfloat[$\beta=0.5$ and $\alpha=0.8$.]{
		\includegraphics[width=0.4\textwidth]{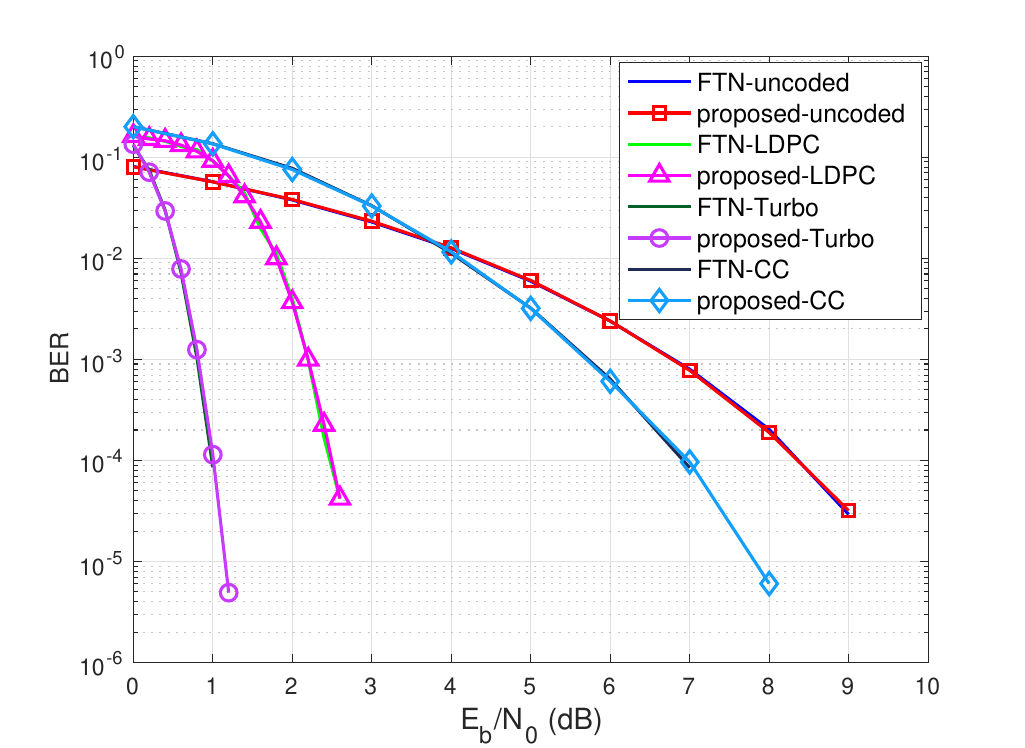}
		\label{fig:ber_0p5_c1}}
	\hspace{0.4in}
	\subfloat[$\beta=0.3$ and $\alpha=0.9$.]{
		\includegraphics[width=0.4\textwidth]{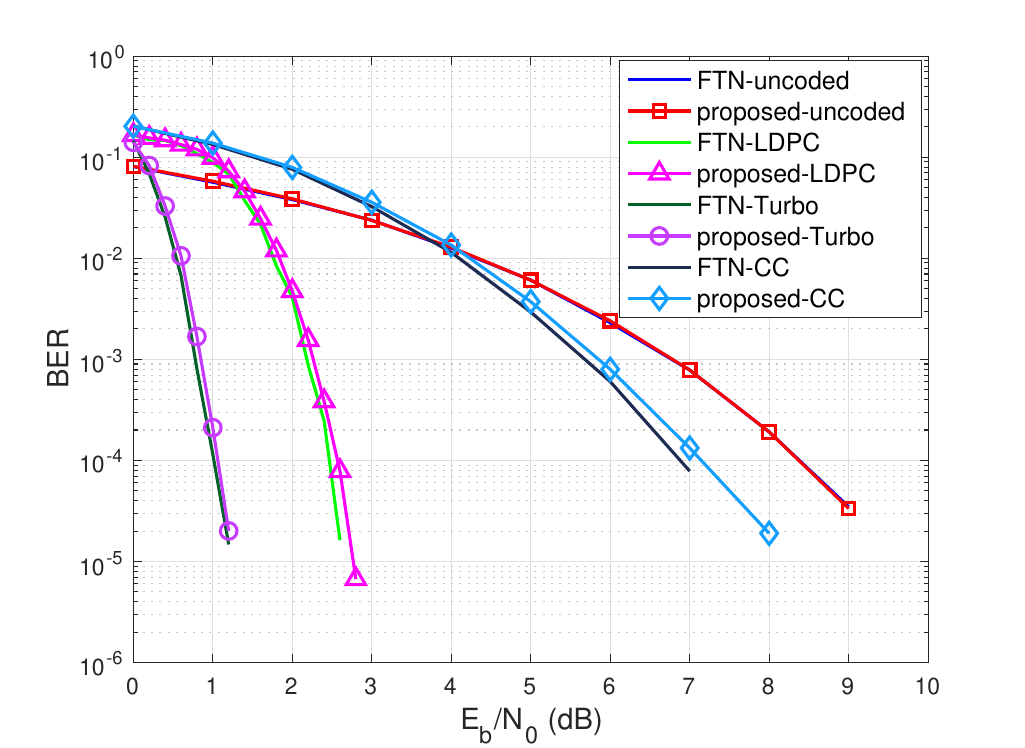}
		\label{fig:ber_0p3_c1}}
	\caption{BER performance of the proposed scheme and the FTN signaling.}
	\label{fig:ber_total}
	\end{figure}

	The simulation results are demonstrated by Fig. \ref{fig:ber_total}\subref{fig:ber_0p5_c1} and Fig. \ref{fig:ber_total}\subref{fig:ber_0p3_c1}.
	
	As seen, with the same SE, the proposed VPR-based secure transmission can achieve nearly the same BER performance as the conventional FTN signaling. It means that the proposed scheme can achieve security at the expense of negligible BER performance degradation.

\subsection{Comparison Between the Simplified Packing Ratio Estimation and the Original Architecture}

In this part, we compare the simplified packing ratio estimation and its original architecture \cite{song2019blind} by the complexity and the accuracy. For the convenience of representation, we only provide the complexity of the branch for analysis on $\alpha=0.7$, while the total complexity is approximately proportional to it. Table \ref{tab:complexity} provides the complexity comparison between these two schemes.

%todo 这个地方的multi-add需要再加一下per estimation

\begin{table*}[!ht]
	\renewcommand\arraystretch{1}
	\centering
	\caption{The complexity comparison between the proposed simplified estimation and its original structure.}
	\label{tab:complexity}
	\begin{tabular}{ccccccccccc} 
		\toprule	
		Algorithm & MUX & DEMUX & sum & max & S/P & $\lVert\mathbf{W}_1\rVert_0$ & $\lVert\mathbf{W}_2\rVert_0$ & $\lVert\mathbf{W}_3\rVert_0$  & $\lVert\mathbf{W}_4\rVert_0$ & multi-add \\
		\midrule
		Original Structure & 1 & 2 & $ 10 $ & 1 & $ 10 $ & 20k & 500k & 250k & 0.25k & 645.25k \\
		Proposed Structure & 0 & 0 & 1 & 0 & 1 & 10k &250k & 62.5K & 0.125k & about 32.263k  \\
		\bottomrule
	\end{tabular}
\end{table*}

The proposed structure nearly removes all the MUX, DEMUX, sum, maximum and S/P operations in the original design. Also, in the sparse DNN employed in our proposed simplified estimation, the number of non-zero weights in each layer has been reduced to half of that in the original network. Significantly, benefiting from the sparse DNN and the single branch structure, the number of multiply-add operations required for each estimation has been reduced to 5\% of that in the original architecture. This allows more flexibility for researchers to balance the resource of time and space in practical implementation.

%\begin{table}[ht]
%	\centering
%	\renewcommand\arraystretch{1.5}
%	\caption{Probability of true decisions for different $\alpha$ and $\alpha_k$ values
%		when $E_b/N_o$ = 4 dB}
%	\label{tab:true_decisions}
%	\begin{tabular}{|c|c|c|c|c|c|}
%		\hline
%		\diagbox{$\alpha$}{$P_{true}$}{$\alpha_k$} & 1 & 0.9 & 0.8 & 0.75 & 0.7 \\
%		\hline
%		1 & 0.7876 & 0.2063 & 0.1572 & 0.1132 & 0.1185 \\
%		\hline
%		0.9 & 0.1513 & 0.7141 & 0.1485 & 0.1297 & 0.1284 \\
%		\hline
%		0.8 & 0.0873 & 0.1316 & 0.6222 & 0.1508 & 0.1821 \\
%		\hline
%		0.75 & 0.0573 & 0.0935 & 0.1532 & 0.5094 & 0.2126 \\
%		\hline
%		0.7 & 0.0386 & 0.0643 & 0.1347 & 0.1709 & 0.5692 \\
%		\hline
%	\end{tabular}
%\end{table}

To more visually demonstrate the performance, we employ the accuracy as \cite{song2019blind}
\begin{equation}\label{eq:accuracy}
\begin{split}
P_{\mathrm{acc}}&= \sum_{m=1}^{M} \sum_{n=0}^{m-1} \left( C_{M}^{m} C_{M}^{n}\left(p_{1}\right)^{m}\left(1-p_{1}\right)^{(M-m)} \left(p_{2}\right)^{n}\left(1-p_{2}\right)^{(M-n)}  \right)
\end{split} ,
\end{equation}
where $M$ is the number of decisions applied to determine the final estimated value of $\alpha$. $p_1$ is the probability that the analysis branch for $\alpha_k=\alpha$ outputs integer $1$ (i.e., the diagonal items in Fig. \ref{fig:heatmap}). And $p_2$ is the maximum probabilities that the analysis branches for $\alpha_k\ne \alpha$ produce integer $1$ (i.e. the maximum one of non-diagonal items within each row in Fig. \ref{fig:heatmap}).

\begin{figure}[ht]
	\centering
	\includegraphics[width=0.35\linewidth]{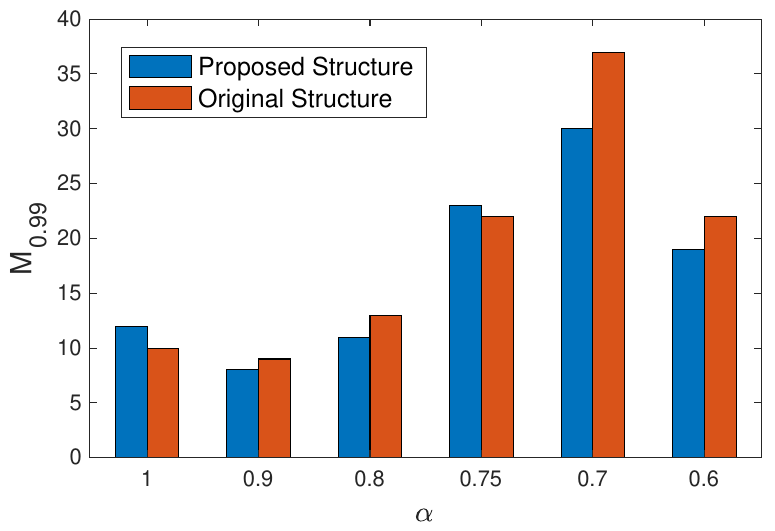}
	\caption{The comparison of the proposed simplified estimation and its original structure in the minimum times of decisions required to achieve a 99\% accuracy. }
	\label{fig:m099}
\end{figure}

Fig. \ref{fig:m099} shows the minimum number of decisions required to achieve a 99\% accuracy ($P_{acc}>0.99$). As seen, the proposed simplified estimation can converge nearly as fast as the original structure within 35 decisions, while the complexity has been greatly reduced.

\subsection{Complexity of the Simplified Estimation and Other Common DL Networks}

Here, two common deep learning networks named Transformer \cite{vaswani2017attention} and Inception-v4 \cite{szegedy2017inception} are considered for the comparison of complexity. They are both proposed by Google and have been widely employed in natural language processing (NLP) and computer vision (CV) research fields. Their effectiveness and complexity have been verified by mass researchers and applications.

The complexity comparison is shown in Table \ref{tab:complexity_comparison}. As seen, the proposed scheme has an obviously lower complexity than the selected widely employed networks.

\subsection{The Robustness of the Simplified Estimation to SNR Values}
Here, the performance of the proposed simplified estimation for $\alpha=0.9$ in AWGN channels under different SNR values is listed in Table \ref{tab:different}. As shown, although the model is trained at SNR=4dB, it can work well for other SNR values. It can effectively reduce the resource required for the proposed estimation during both the training and the implementation stages. 

\begin{table}
	\captionsetup{type=table}
	\caption{Complexity comparison between the proposed scheme and some common networks.}
	\centering
	\setlength{\belowcaptionskip}{0.1cm} %段后
	\renewcommand\arraystretch{1}
	\setlength{\tabcolsep}{3mm}
	\label{tab:complexity_comparison}

	\begin{tabular}[c]{|c|c|c|c|c|}
		\hline
		Network & Proposed & Transformer (base) & Transformer (big) & Resnet \\
		\hline
	    Parameters & 0.645M & 65M & 213M & 48M \\
	    \hline
	    FLOPs & $6.5\times 10^4$ & $3.3\times 10^{18}$ & $2.3 \times 10^{19}$ & $1.3\times10^{10}$ \\
		\hline
	\end{tabular}
\end{table}

%todo 原来是1.6和4.5
\begin{table}[ht]
	\centering
	\renewcommand\arraystretch{1}
	\setlength{\tabcolsep}{4mm}
	\caption{Performance of the proposed estimation which is trained at SNR=4dB and tested for different SNRs.}
	\label{tab:different}
	
\begin{tabular}{|c|c|c|c|c|}
	\hline
	SNR & 4dB & 3dB & 2dB & 1dB \\
	\hline
	$p_1$ & 0.7534 & 0.5084 & 0.4372 & 0.2914 \\
	\hline
	$p_2$ & 0.1632 & 0.1596 & 0.1463 & 0.1485  \\
	\hline
	$M_{0.99}$ & 8 & 21 & 28 & 95 \\
	\hline
	$M_{0.999}$ & 13 & 35 & 47 & 163 \\
	\hline
\end{tabular}
%\begin{tabular}{c|c|c}
%	\hline 
%	$E_b/N_o$ & $p_1$ & $p_2$ \\ 
%	\hline 
%	4dB & 0.7876 & 0.2063 \\ 
%	\hline 
%	3dB & 0.5084 & 0.1596 \\ 
%	\hline 
%	2dB & 0.4372 & 0.1463 \\ 
%	\hline 
%	1dB & 0.2914 & 0.1485 \\ 
%	\hline 
%\end{tabular} 
\end{table}

\section{Conclusion} \label{sec:conclusion}
This paper proposed intelligent VPR transmissions for high SE and security, respectively, based on FTN and DL. The VPR-based system achieved a higher SE without consuming extra spectrum resources and modifying the existing communication paradigms (e.g., spectrum allocation or frame structure). Also, considering security, a dynamic generation scheme was proposed to produce secret and randomly distributed positions for the segments of the VPR system. The scheme was demonstrated to be effective in avoiding detection and attack. In addition, we derived the closed-form expression for the capacity of the proposed VPR system in different channels, which were also effective for conventional FTN signaling. Finally, a simplified symbol packing ratio, which had been employed in the proposed system, was developed in this paper. Simulation results proved that it achieved nearly the same performance as the original structure with only 5\% of the complexity in the original design.

In fact, there are still many open issues with the proposed VPR system beckoning further research. For example, how to design an effective switching strategy for the VPR system considering practical factors (e.g., interference, relay, energy harvesting, etc.), especially the nondeterministic polynomial (NP)-hard scenario is considered? How to derive the closed-form SE of the proposed scheme in other channels? Is it possible to develop a better packing ratio estimation algorithm to further improve the robustness of the VPR system? These issues will be studied in our future works.

%The receiver can estimate the current $\alpha$ by the estimation architecture and the shared information of the starting position. Then, the link turns to be a conventional FTN signaling. But the eavesdropper will suffer from severe performance degradation because of the absence of the required information. 


%\section*{Acknowledgment}
%This work is supported in part by the National Natural Science Foundation of China (62001354), the National Natural Science Foundation of China (61901325), and Natural Sciences and Engineering Research Council of Canada (NSERC).



\bibliographystyle{IEEEtran}
\bibliography{database}

\begin{thebibliography}{10}
\providecommand{\url}[1]{#1}
\csname url@samestyle\endcsname
\providecommand{\newblock}{\relax}
\providecommand{\bibinfo}[2]{#2}
\providecommand{\BIBentrySTDinterwordspacing}{\spaceskip=0pt\relax}
\providecommand{\BIBentryALTinterwordstretchfactor}{4}
\providecommand{\BIBentryALTinterwordspacing}{\spaceskip=\fontdimen2\font plus
\BIBentryALTinterwordstretchfactor\fontdimen3\font minus
  \fontdimen4\font\relax}
\providecommand{\BIBforeignlanguage}[2]{{%
\expandafter\ifx\csname l@#1\endcsname\relax
\typeout{** WARNING: IEEEtran.bst: No hyphenation pattern has been}%
\typeout{** loaded for the language `#1'. Using the pattern for}%
\typeout{** the default language instead.}%
\else
\language=\csname l@#1\endcsname
\fi
#2}}
\providecommand{\BIBdecl}{\relax}
\BIBdecl

\bibitem{mazo1975faster}
J.~E. Mazo, ``Faster-than-{Nyquist} signaling,'' \emph{Bell Syst. Technical
  J.}, vol.~54, no.~8, pp. 1451--1462, 1975.

\bibitem{liveris2003exploiting}
A.~D. Liveris and C.~N. Georghiades, ``Exploiting faster-than-{Nyquist}
  signaling,'' \emph{IEEE Trans. Commun.}, vol.~51, no.~9, pp. 1502--1511,
  2003.

\bibitem{anderson2009new}
J.~B. Anderson, A.~Prlja, and F.~Rusek, ``New reduced state space {BCJR}
  algorithms for the {ISI} channel,'' in \emph{Proc. IEEE Int. Symp. Inf.
  Theory, Seoul, South Korea}.\hskip 1em plus 0.5em minus 0.4em\relax IEEE,
  2009, pp. 889--893.

\bibitem{bedeer2017very}
E.~Bedeer, M.~H. Ahmed, and H.~Yanikomeroglu, ``A very low complexity
  successive symbol-by-symbol sequence estimator for faster-than-{Nyquist}
  signaling,'' \emph{IEEE Access}, vol.~5, no.~99, pp. 7414--7422, 2017.

\bibitem{song2020dl}
P.~Song, F.~Gong, Q.~Li, G.~Li, and H.~Ding, ``Receiver design for
  faster-than-{Nyquist} signaling: Deep-learning-based architectures,''
  \emph{IEEE Access}, vol.~8, pp. 68\,866--68\,873, 2020.

\bibitem{liu2021novel}
B.~Liu, S.~Li, Y.~Xie, and J.~Yuan, ``A novel sum-product detection algorithm
  for faster-than-nyquist signaling: A deep learning approach,'' \emph{IEEE
  Trans. Commun.}, vol.~69, no.~9, pp. 5975--5987, 2021.

\bibitem{petitpied2021circular}
T.~Petitpied, R.~Tajan, P.~Chevalier, S.~Traverso, and G.~Ferr{\'e}, ``Circular
  faster-than-{Nyquist} signaling for high spectral efficiencies: optimized
  {EP}-based receivers,'' \emph{IEEE Trans. Commun.}, vol.~69, no.~8, pp.
  5487--5501, 2021.

\bibitem{ibrahim2021novel}
A.~Ibrahim, E.~Bedeer, and H.~Yanikomeroglu, ``A novel low complexity
  faster-than-nyquist signaling detector based on the primal-dual
  predictor-corrector interior point method,'' \emph{IEEE Commun. Lett.},
  vol.~25, no.~7, pp. 2370--2374, 2021.

\bibitem{sugiura2013frequency}
S.~Sugiura, ``Frequency-domain equalization of faster-than-{Nyquist}
  signaling,'' \emph{IEEE Wireless Commun. Lett.}, vol.~2, no.~5, pp. 555--558,
  2013.

\bibitem{ishihara2016frequency}
T.~Ishihara and S.~Sugiura, ``Frequency-domain equalization aided iterative
  detection of faster-than-{Nyquist} signaling with noise whitening,'' in
  \emph{Proc. IEEE Int. Conf. Commun. (ICC), Kuala Lumpur, Malaysia}.\hskip 1em
  plus 0.5em minus 0.4em\relax IEEE, 2016, pp. 1--6.

\bibitem{rusek2009multistream}
F.~Rusek and J.~B. Anderson, ``Multistream faster than nyquist signaling,''
  \emph{IEEE Trans. Commun.}, vol.~57, no.~5, pp. 1329--1340, 2009.

\bibitem{che2021multicarrier}
H.~Che, K.~Zhu, and Y.~Bai, ``Multicarrier {faster-than-Nyquist} based on
  efficient implementation and probabilistic shaping,'' \emph{IEEE Access},
  vol.~9, pp. 63\,943--63\,951, 2021.

\bibitem{ishihara2022reduced}
T.~Ishihara and S.~Sugiura, ``Reduced-complexity {FFT}-spread multi-carrier
  faster-than-{Nyquist} signaling in frequency-selective fading channel,''
  \emph{IEEE Open J. Commun. Soc.}, 2022.

\bibitem{ma2021generalized}
Y.~Ma, N.~Wu, A.~Zhang, B.~Li, and L.~Hanzo, ``Generalized approximated message
  passing equalization for multi-carrier faster-than-{Nyquist} signaling,''
  \emph{IEEE Trans. Veh. Technol.}, 2021.

\bibitem{anderson2006improving}
J.~B. Anderson and F.~Rusek, ``Improving {OFDM}: Multistream
  faster-than-{Nyquist} signaling,'' in \emph{proc, 4th Int. Symp. Turbo Codes
  \& Related Topics; 6th Int. ITG-Conference on Source and Channel
  Coding}.\hskip 1em plus 0.5em minus 0.4em\relax VDE, 2006, pp. 1--5.

\bibitem{rusek2009existence}
F.~Rusek, ``On the existence of the {Mazo-limit} on {MIMO} channels,''
  \emph{IEEE Trans. Wireless Commun.}, vol.~8, no.~3, pp. 1118--1121, 2009.

\bibitem{abebe2018ftn}
A.~T. Abebe and C.~G. Kang, ``{FTN}-based {MIMO} transmission as a {NOMA}
  scheme for efficient coexistence of broadband and sporadic traffics,'' in
  \emph{proc, 2018 IEEE 87th Veh. Technol. Conf. (VTC Spring)}.\hskip 1em plus
  0.5em minus 0.4em\relax IEEE, 2018, pp. 1--5.

\bibitem{yuhas2015capacity}
M.~Yuhas, Y.~Feng, and J.~Bajcsy, ``On the capacity of faster-than-{Nyquist}
  {MIMO} transmission with {CSI} at the receiver,'' in \emph{proc, 2015 IEEE
  Globecom Workshops (GC Wkshps)}.\hskip 1em plus 0.5em minus 0.4em\relax IEEE,
  2015, pp. 1--6.

\bibitem{mcguire2016faster}
M.~McGuire, A.~Dimopoulos, and M.~Sima, ``Faster-than-{Nyquist} single-carrier
  {MIMO} signaling,'' in \emph{proc, 2016 IEEE Globecom Workshops (GC
  Wkshps)}.\hskip 1em plus 0.5em minus 0.4em\relax IEEE, 2016, pp. 1--7.

\bibitem{wen2022joint}
S.~Wen, G.~Liu, C.~Liu, H.~Qu, L.~Zhang, and M.~A. Imran, ``Joint precoding and
  pre-equalization for {faster-Than-Nyquist} transmission over multipath fading
  channels,'' \emph{IEEE Trans. Veh. Technol.}, 2022.

\bibitem{ishihara2017iterative}
T.~Ishihara and S.~Sugiura, ``Iterative frequency-domain joint channel
  estimation and data detection of {faster-than-Nyquist} signaling,''
  \emph{IEEE Trans. Wireless Commun.}, vol.~16, no.~9, pp. 6221--6231, 2017.

\bibitem{li2020joint}
Q.~Li, F.-K. Gong, P.-Y. Song, G.~Li, and S.-H. Zhai, ``Joint channel
  estimation and precoding for {faster-than-Nyquist} signaling,'' \emph{IEEE
  Trans. Veh. Technol.}, vol.~69, no.~11, pp. 13\,139--13\,147, 2020.

\bibitem{ishihara2021evolution}
T.~Ishihara, S.~Sugiura, and L.~Hanzo, ``The evolution of faster-than-{Nyquist}
  signaling,'' \emph{IEEE Access}, vol.~9, 2021.

\bibitem{zhou2019digital}
J.~Zhou, M.~Guo, Y.~Qiao, H.~Wang, L.~Liu \emph{et~al.}, ``Digital signal
  processing for faster-than-{Nyquist} non-orthogonal systems: An overview,''
  in \emph{proc, 2019 26th Int. Conf. Telecommun. (ICT)}.\hskip 1em plus 0.5em
  minus 0.4em\relax IEEE, 2019, pp. 295--299.

\bibitem{fan2017faster}
J.~Fan, S.~Guo, X.~Zhou, Y.~Ren, G.~Y. Li, and X.~Chen,
  ``{Faster-than-{Nyquist} signaling}: an overview,'' \emph{IEEE Access},
  vol.~5, pp. 1925--1940, 2017.

\bibitem{rusek2009constrained}
F.~Rusek and J.~B. Anderson, ``Constrained capacities for faster-than-{Nyquist}
  signaling,'' \emph{IEEE Trans. Inf. Theory}, vol.~55, no.~2, pp. 764--775,
  2009.

\bibitem{wang2018hopping}
J.~Wang, W.~Tang, X.~Li, and S.~Li, ``Filter hopping based
  faster-than-{Nyquist} signaling for physical layer security,'' \emph{IEEE
  Wireless Commun. Lett.}, vol.~64, no.~5, pp. 2122--2128, 2018.

\bibitem{dvb_standard}
ETSI, ``Digital video broadcasting ({DVB}); implementation guidelines for the
  second generation system for broadcasting, interactive services, news
  gathering and other broadband satellite applications; part 1: {DVB-S2},''
  available:
  \url{https://www.etsi.org/deliver/etsi_tr/102300_102399/10237601/01.02.01_60/tr_10237601v010201p.pdf},
  2015.

\bibitem{aono2005wireless}
T.~Aono, K.~Higuchi, T.~Ohira, B.~Komiyama, and H.~Sasaoka, ``Wireless secret
  key generation exploiting reactance-domain scalar response of multipath
  fading channels,'' \emph{IEEE Trans. Antennas Propag.}, vol.~53, no.~11, pp.
  3776--3784, 2005.

\bibitem{mathur2008radio}
S.~Mathur, W.~Trappe, N.~Mandayam, C.~Ye, and A.~Reznik, ``Radio-telepathy:
  extracting a secret key from an unauthenticated wireless channel,'' in
  \emph{Proceedings of the 14th ACM international conference on Mobile
  computing and networking}, 2008, pp. 128--139.

\bibitem{patwari2009high}
N.~Patwari, J.~Croft, S.~Jana, and S.~K. Kasera, ``High-rate uncorrelated bit
  extraction for shared secret key generation from channel measurements,''
  \emph{IEEE Trans. Mob. Comput.}, vol.~9, no.~1, pp. 17--30, 2009.

\bibitem{brassard1993secret}
G.~Brassard and L.~Salvail, ``Secret-key reconciliation by public discussion,''
  in \emph{proc, Workshop on the Theory and Application of of Cryptographic
  Techniques}.\hskip 1em plus 0.5em minus 0.4em\relax Springer, 1993, pp.
  410--423.

\bibitem{rukhin2001statistical}
A.~Rukhin, J.~Soto, J.~Nechvatal, M.~Smid, and E.~Barker, ``A statistical test
  suite for random and pseudorandom number generators for cryptographic
  applications,'' Booz-allen and hamilton inc mclean va, Tech. Rep., 2001.

\bibitem{madiseh2012applying}
M.~G. Madiseh, S.~W. Neville \emph{et~al.}, ``Applying beamforming to address
  temporal correlation in wireless channel characterization-based secret key
  generation,'' \emph{IEEE Trans. Inf. Forensics Secur.}, vol.~7, no.~4, pp.
  1278--1287, 2012.

\bibitem{aldaghri2020physical}
N.~Aldaghri and H.~Mahdavifar, ``Physical layer secret key generation in static
  environments,'' \emph{IEEE Trans. Inf. Forensics Secur.}, vol.~15, pp.
  2692--2705, 2020.

\bibitem{sayeed2008secure}
A.~Sayeed and A.~Perrig, ``Secure wireless communications: Secret keys through
  multipath,'' in \emph{proc, 2008 IEEE International Conference on Acoustics,
  Speech and Signal Processing}.\hskip 1em plus 0.5em minus 0.4em\relax IEEE,
  2008, pp. 3013--3016.

\bibitem{goodfellow2016deep}
I.~Goodfellow, Y.~Bengio, and A.~Courville, \emph{Deep Learning}.\hskip 1em
  plus 0.5em minus 0.4em\relax MIT press, 2016.

\bibitem{song2019blind}
P.~Song, F.~Gong, and Q.~Li, ``Blind symbol packing ratio estimation for
  faster-than-nyquist signalling based on deep learning,'' \emph{Electron.
  Lett.}, vol.~55, no.~21, pp. 1155--1157, 2019.

\bibitem{rice1944mathematical}
S.~O. Rice, ``Mathematical analysis of random noise,'' \emph{Bell Syst.
  Technical J.}, vol.~23, no.~3, pp. 282--332, 1944.

\bibitem{nakagami1960m}
M.~Nakagami, ``The m-distribution-a general formula of intensity distribution
  of rapid fading,'' \emph{Statal Methods in Radio Wave Propagation}, pp. 3--6,
  1960.

\bibitem{cubukcu2012root}
E.~Cubukcu, ``Root raised cosine (rrc) filters and pulse shaping in
  communication systems,'' in \emph{proc, AIAA Conference}, no. JSC-CN-26387,
  2012.

\bibitem{abramowitz1964handbook}
M.~Abramowitz and I.~A. Stegun, \emph{Handbook of mathematical functions with
  formulas, graphs, and mathematical tables}.\hskip 1em plus 0.5em minus
  0.4em\relax US Government printing office, 1964, vol.~55.

\bibitem{thomas1961calculus}
G.~B. Thomas and R.~L. Finney, \emph{Calculus}.\hskip 1em plus 0.5em minus
  0.4em\relax Addison-Wesley Publishing Company, 1961.

\bibitem{ross2004differential}
C.~C. Ross, \emph{Differential equations: an introduction with
  Mathematica}.\hskip 1em plus 0.5em minus 0.4em\relax Springer Science \&
  Business Media, 2004.

\bibitem{flajolet1995mellin}
P.~Flajolet, X.~Gourdon, and P.~Dumas, ``Mellin transforms and asymptotics:
  Harmonic sums,'' \emph{Theoretical Comput. Sci.}, vol. 144, no. 1-2, pp.
  3--58, 1995.

\bibitem{edition2007table}
S.~Edition, ``Table of integrals, series, and products,'' 2007.

\bibitem{bateman1953higher}
H.~Bateman, \emph{Higher transcendental functions [volumes i-iii]}.\hskip 1em
  plus 0.5em minus 0.4em\relax McGraw-Hill Book Company, 1953, vol.~1.

\bibitem{li2018reduced}
S.~Li, B.~Bai, J.~Zhou, P.~Chen, and Z.~Yu, ``Reduced-complexity equalization
  for faster-than-{Nyquist} signaling: New methods based on {Ungerboeck}
  observation model,'' \emph{IEEE Trans. Commun.}, vol.~66, no.~3, pp.
  1190--1204, 2018.

\bibitem{9502043}
``{IEEE} standard for information technology--telecommunications and
  information exchange between systems - local and metropolitan area
  networks--specific requirements - part 11: Wireless lan medium access control
  ({MAC}) and physical layer ({PHY}) specifications - redline,'' \emph{IEEE Std
  802.11-2020 (Revision of IEEE Std 802.11-2016) - Redline}, pp. 1--7524, 2021.

\bibitem{vaswani2017attention}
A.~Vaswani, N.~Shazeer, N.~Parmar, J.~Uszkoreit, L.~Jones, A.~N. Gomez,
  {\L}.~Kaiser, and I.~Polosukhin, ``Attention is all you need,''
  \emph{Advances in neural information processing systems}, vol.~30, 2017.

\bibitem{szegedy2017inception}
C.~Szegedy, S.~Ioffe, V.~Vanhoucke, and A.~A. Alemi, ``Inception-v4,
  inception-resnet and the impact of residual connections on learning,'' in
  \emph{proc, Thirty-first AAAI conference on artificial intelligence}, 2017.

\end{thebibliography}

\end{document}